\PassOptionsToPackage{dvipsnames}{xcolor}
\documentclass{lmcs}
\pdfoutput=1

\usepackage{lastpage}
\lmcsdoi{15}{2}{5}
\lmcsheading{}{\pageref{LastPage}}{}{}%
{Oct.~31,~2018}{Apr.~30,~2019}{}

\usepackage{macros}
\newcommand{\STEP}{
  \begin{array}{@{}l@{}}
\StepRule: \\
  \logicrule{
\begin{array}{@{}l@{}}
\models \varphi \ra \bigvee_{\lhs \Ra \rhs \ \in \ \Sem} \exists
\FV{\lhs} \ \lhs \\
\models \exists c \ (\varphi[c/\square] \andx \lhs[c/\square]) \andx \rhs \ra
\varphi' \qquad \mbox{ for each } \ \lhs \Ra \rhs \ \in \ \Sem \\
\end{array}}{\sequent[\cal C]{\cal A}{\varphi}{\varphi'}}
\end{array}}

\newcommand{\AXIOM}{
  \begin{array}{@{}l@{}}
  \AxiomRule: \\[1ex]
  \logicrule{
\varphi \Rra \varphi' \ \ \in \ {\cal A}
}
{\sequent[\cal C]{\cal A}{\varphi}{\varphi'}}
\end{array}
}

\newcommand{\REFLEXIVITY}{
  \begin{array}{@{}l@{}}
  \ReflexivityRule: \\[1ex]
  \logicrule{\cdot}{
\sequent{\cal A}{\varphi}{\varphi}
}
  \end{array}
}

\newcommand{\TRANSITIVITY}{
\begin{array}{@{}l@{}}
\TransitivityRule: \\[1ex]
\logicrule{
\sequent[\cal C]{\cal A}{\varphi_1}{\varphi_2} \qquad
\sequent{{\cal A} \cup {\cal C}}{\varphi_2}{\varphi_3}
}{\sequent[\cal C]{\cal A}{\varphi_1}{\varphi_3}} \\
\end{array}
}

\newcommand{\CASE}{
\begin{array}{@{}l@{}}
\CaseAnalysisRule: \\[1ex]
\logicrule{
\sequent[\cal C]{\cal A}{\varphi_1}{\varphi} \qquad
\sequent[\cal C]{\cal A}{\varphi_2}{\varphi}
}
{\sequent[\cal C]{\cal A}{\varphi_1 \orx \varphi_2}{\varphi}} \\
\end{array}
}

\newcommand{\CONSEQUENCE}{
\begin{array}{@{}l@{}}
\ConsequenceRule: \\[1ex]
\logicrule{
\models\varphi_1\ra\varphi_1' \qquad \
\sequent[\cal C]{\cal A}{\varphi_1'}{\varphi_2'} \qquad
\models\varphi_2'\ra\varphi_2
}{\sequent[\cal C]{\cal A}{\varphi_1}{\varphi_2}}
\end{array}
}

\newcommand{\ABSTRACTION}{
\begin{array}{@{}l@{}}
\AbstractionRule: \\[1ex]
\logicrule{
\sequent[\cal C]{\cal A}{\varphi}{\varphi'} \qquad X \cap \FV{\varphi'} = \emptyset
}{\sequent[\cal C]{\cal A}{\exists X \ \varphi}{\varphi'}}
\end{array}
}

\newcommand{\CIRCULARITYSANSNAME}{
\logicrule{
\sequent[{\cal C} \cup \{ \varphi \Rra \varphi' \}]{\cal A}{\varphi}{\varphi'}
}{\sequent[\cal C]{{\cal A}}{\varphi}{\varphi'}}
}

\newcommand{\CIRCULARITY}{
\begin{array}{@{}l@{}}
\CircularityRule: \\[1ex]
\CIRCULARITYSANSNAME
\end{array}
}

\newcommand{\PROOFSYSTEM}{
\[\begin{array}{@{}l@{\hspace*{-15ex}}r@{}}
\STEP & \AXIOM \\[5ex]
\REFLEXIVITY & \TRANSITIVITY \\[5ex]
\CASE & \ABSTRACTION \\[5ex]
\CONSEQUENCE & \CIRCULARITY \\
\end{array}\]
}

\makeatletter
\DeclareRobustCommand*\cal{\@fontswitch\relax\mathcal}
\makeatother
\begin{document}


\title{All-Path Reachability Logic}

\author[\Stefanescu]{Andrei \Stefanescu\rsuper{{a,*}}}

\address{\lsuper{a}Galois, Inc., USA}
\email{andrei@galois.com}
\address{\lsuper{b}Alexandru Ioan Cuza University, Romania}
\email{\{stefan.ciobaca,radu.mereuta\}@info.uaic.ro}
\address{\lsuper{c}Runtime Verification, Inc., USA}
\email{brandon.moore@runtimeverification.com}
\address{\lsuper{d}University of Illinois at Urbana-Champaign, USA}
\email{grosu@illinois.edu}
\thanks{\lsuper{*}Most of the present work was done while still being PhD
  students at the University of Illinois at Urbana-Champaign.}
\address{\lsuper{e}University of Bucharest, Romania}
\email{traian.serbanuta@fmi.unibuc.ro}
\address{\vspace{-2cm}}

\author[\Ciobaca]{Stefan \Ciobaca\rsuper{b}}

\author[Mereuta]{Radu Mereuta\rsuper{{b,c}}}

\author[Moore]{Brandon M. Moore\rsuper{{c,*}}}

\author[\Rosu]{Grigore \Rosu\rsuper{{d,c}}}

\author[Florin]{Traian Florin \Serbanuta\rsuper{{c,e}}}

\maketitle

\address{\vspace{-2mm}}
\begin{abstract}
This paper presents a language-independent proof system for reachability
properties of programs written in non-deterministic (e.g., concurrent) languages,
referred to as {\em all-path reachability logic}.
It derives partial-correctness properties with all-path semantics (a state
satisfying a given precondition reaches states satisfying a given postcondition
on all terminating execution paths).
The proof system takes as axioms any unconditional operational semantics, and is
sound (partially correct) and (relatively) complete, independent of the
object language.
The soundness has also been mechanized in Coq.
This approach is implemented in a tool for semantics-based verification as part
of the \K framework (\url{http://kframework.org}).
%
\end{abstract}
\address{\vspace{-2.5mm}}

\section{Introduction}\label{sec:introduction}


Operational semantics are easy to define and understand.
Giving a language an operational semantics can be regarded as ``implementing'' a
formal interpreter.
Operational semantics require little formal training, scale up well and, being
executable, can be tested.
Thus, operational semantics are typically used as trusted reference models
for the defined languages.
Despite these advantages, operational semantics are rarely used directly for
program verification (i.e., verifying properties of a given program, rather than
performing meta-reasoning about a given language),
because such proofs tend to be low-level and tedious, as they
involve formalizing and working directly with the corresponding transition
system.
Hoare or dynamic logics allow higher level reasoning at the cost of (re)defining
the language as a set of abstract proof rules, which are harder to understand
and trust.
The state-of-the-art in mechanical program verification is to develop and prove
such language-specific proof systems sound w.r.t to a trusted operational
semantics~\cite{Nipkow-FAC98,DBLP:journals/jlp/Jacobs04,DBLP:conf/esop/Appel11},
but that needs to be done for each language separately.

Defining more semantics for the same language and proving the soundness of
one semantics in terms of another are highly uneconomical tasks where real
programming languages are concerned, often taking several years to complete.
Ideally, we would like to have only one semantics for a language, together with
a generic theory and a set of generic tools and techniques allowing us to get all
the benefits of any other semantics without paying the price of defining other
semantics.
Recent
work~\cite{rosu-stefanescu-2012-oopsla,rosu-stefanescu-2012-fm,rosu-stefanescu-2012-icalp,rosu-stefanescu-ciobaca-moore-2013-lics}
shows this {\em is possible}, by proposing a {\em language-independent proof
system} which derives program properties directly from an operational semantics,
{\em at the same proof granularity and compositionality} as a language-specific
axiomatic semantics.
%
%
%
Specifically, it introduces \emph{(one-path) reachability rules}, which
generalize both operational semantics reduction rules and Hoare triples, and
gives a proof system which derives new reachability rules (program properties)
from a set of given reachability rules (the language semantics).

However, the existing proof system has a major limitation: it only derives
reachability rules with a \emph{one-path} semantics, that is, it guarantees a
program property holds on one but not necessarily all execution paths, which
suffices for deterministic languages but not
for non-deterministic (concurrent) languages.
Here we remove this limitation, proposing the first generic all-path reachability proof
system for program verification.


Using {\em matching logic}~\cite{rosu-ellison-schulte-2010-amast} as a
configuration specification formalism (Section~\ref{sec:matching-logic}), where
a pattern $\varphi$ specifies all program configurations that match it, we first
introduce the novel notion of an {\em all-path reachability rule} $\varphi \Rra
\varphi'$ (Section~\ref{sec:reachability-rules}), where $\varphi$ and
$\varphi'$ are matching logic patterns.
Rule $\varphi \Rra \varphi'$ is valid iff any program configuration satisfying
$\varphi$ reaches, on any complete execution path, some configuration
satisfying $\varphi'$.
This subsumes Hoare-style partial correctness in non-deterministic languages.
%
We then present a proof system for deriving an all-path reachability rule
$\varphi \ral \varphi'$ from a set $\Sem$ of semantics rules
(Section~\ref{sec:proof-system}).
$\Sem$ consists of reduction rules $\varphi_l \Ra \varphi_r$, where $\varphi_l$
and $\varphi_r$ are simple patterns as encountered in operational semantics
(Section~\ref{sec:IMP}), which can be non-deterministic.
The proof system derives more general sequents
``$\sequent[\cal C]{\cal A}{\varphi}{\varphi'}$'', with ${\cal A}$ and ${\cal C}$
two sets of reachability rules.
Intuitively, ${\cal A}$'s rules ({\em axioms}) are already established valid,
and thus can be immediately used.
Those in ${\cal C}$ ({\em circularities}) are only claimed valid, and can be used
only after taking execution steps based on the rules in $\Sem$ or ${\cal A}$.
The most important proof rules are
\begin{align*}
& \STEP \\
& \CIRCULARITY
\end{align*}
\Step is the key proof rule which deals with non-determinism: it derives a
sequent where $\varphi$ reaches $\varphi'$ in one step on all paths.
The first premise ensures that any configuration satisfying $\varphi$ has
successors, the second that all successors satisfy $\varphi'$ ($\square$ is
the configuration placeholder).
\Circularity adds the current goal to ${\cal C}$ at any point in a proof, and
generalizes the various language-specific axiomatic
semantics invariant rules in a language-independent way
(this form was introduced in~\cite{rosu-stefanescu-2012-oopsla}).

We illustrate on examples how our proof system enables state exploration
(similar to symbolic model-checking), and verification of program properties
(Section~\ref{sec:IMP}).
We show that our proof system is sound
(Section~\ref{sec:soundness}) and relatively complete
(Section~\ref{sec:completeness}).
We describe our implementation of the proof system as part of the \K
framework~\cite{rosu-serbanuta-2010-jlap} (Section~\ref{sec:implementation}).

\paragraph{\textbf{Contributions}} This paper makes the following specific
contributions:
\begin{enumerate}

\item A language-independent proof system for deriving all-path reachability
  properties, with (complete) proofs of its soundness and relative completeness;
  the soundness result has also been mechanized in Coq, to serve as a foundation
  for certifiable verification.

\item An implementation of it as part of the \K framework (\url{http://kframework.org}).

\end{enumerate}

An earlier, shorter version of this paper appeared at RTA-TCLA 2014~\cite{stefanescu-ciobaca-mereuta-moore-serbanuta-rosu-2014-rta}.
The main differences are the inclusion of proofs for soundness
(Section~\ref{sec:soundness}), and relative completeness
(Section~\ref{sec:completeness}).

\section{Topmost Matching Logic}\label{sec:matching-logic}

Matching logic~\cite{rosu-2017-lmcs,rosu-2015-rta} is a first-order logic (FOL) variant
designed for specifying and reasoning about structure by means of patterns and
pattern matching.
In this section, we briefly recall topmost matching
logic~\cite{rosu-ellison-schulte-2010-amast}, a subset of the full matching
logic theory described in~\cite{rosu-2017-lmcs,rosu-2015-rta}.
We employ topmost matching logic to specify and reason about arbitrary program
configurations.
For simplicity, we use ``matching logic'' instead of ``topmost matching logic''
in this paper.
A matching logic formula, called a {\em pattern}, is a first-order logic (FOL)
formula with special predicates, called basic patterns.
A {\em basic pattern} is a configuration term with variables.
Intuitively, a pattern specifies both structural and logical
constraints: a configuration satisfies the pattern iff it matches the structure
(basic patterns)
and satisfies the constraints. 

Matching logic is parametric in a signature and a model of configurations,
making it a prime candidate for expressing state properties in a
language-independent verification framework. 
The configuration signature can be as simple as that of \IMP
(Fig.~\ref{fig:IMP}), or as complex as that of the C
language~\cite{ellison-rosu-2012-popl,hathhorn-ellison-rosu-2015-pldi}
(with more than 100 semantic components).

We use basic concepts from multi-sorted first-order logic.
Given a {\em signature} $\Sigma$ which specifies the sorts and arities of the
function symbols (constructors or operators) used in configurations, let
$T_\Sigma(\Var)$ denote the {\em free $\Sigma$-algebra} of terms with variables
in $\Var$.
$T_{\Sigma,s}(\Var)$ is the set of $\Sigma$-terms of sort $s$.
A valuation $\rho \!:\! \Var \!\ra\! {\cal T}$ with $\cal T$ a $\Sigma$-algebra
extends uniquely to a (homonymous) {\em $\Sigma$-algebra morphism} $\rho \!:\!
T_\Sigma(\Var) \!\ra\! {\cal T}$.
Many mathematical structures needed for language semantics have been defined as
$\Sigma$-algebras, including: boolean algebras, natural/integer/rational
numbers, lists, sets, bags (or multisets), maps (e.g., for states, heaps),
trees, queues, stacks, etc.

Let us fix the following:
(1) an algebraic signature $\Sigma$, associated to some desired configuration
syntax, with a distinguished sort $\Cfg$,
(2) a sort-wise infinite set $\Var$ of variables, and
(3) a $\Sigma$-algebra $\cal T$, the {\em configuration model}, which may but
need not be a term algebra.
As usual, $\TCfg$ denotes the elements of $\cal T$ of sort $\Cfg$,
called {\em configurations}.

\begin{defiC}[\cite{rosu-ellison-schulte-2010-amast}]
A matching logic formula, or a \textbf{pattern}, is a first-order logic (FOL)
formula which additionally allows terms in $T_{\Sigma,\Cfg}(\Var)$,
called \textbf{basic patterns}, as predicates.
A pattern is \textbf{structureless} if it contains no basic patterns.

We define satisfaction $\mlmodels{\gamma}{\rho}{\varphi}$ over
configurations $\gamma \!\in\! {\TCfg}$, valuations $\rho \!:\! \Var \!\ra\! {\cal
T}$ and patterns $\varphi$ as follows (among the FOL constructs, we only
show~$\exists$):
\[\begin{array}{l}
\mlmodels{\gamma}{\rho}{\exists X\ \varphi} \ \ \mbox{iff} \ \
\mlmodels{\gamma}{\rho'}{\varphi} \mbox{ \ for some $\rho'\!:\!\Var \!\ra\!
{\cal T}$}
\ \mbox{with $\rho'(y)=\rho(y)$ for all $y\in\Var \backslash X$ }
\\[-.3ex]
\!\colorbox{Grey}{\!$\begin{array}{@{}lcl} \mlmodels{\gamma}{\rho}{\pi} & \,
\mbox{\ \ \ \ \ \, iff} & \gamma = \rho(\pi) \end{array}$ \!\!\!} \mbox{\ \ \ \
where $\pi\in T_{\Sigma,\Cfg}(\Var)$} \\
\end{array}\]
We write $\models\varphi$ when $\mlmodels{\gamma}{\rho}{\varphi}$ for all
$\gamma\in {\cal T}_{\Cfg}$ and all $\rho:\Var\ra{\cal T}$.
\end{defiC}

A basic pattern $\pi$ is satisfied by all the configurations $\gamma$ that
\emph{match} it; in $\mlmodels{\gamma}{\rho}{\!\pi}$ the $\rho$ can be thought
of as the ``witness'' of the matching, and can be further constrained in a
pattern.
For instance, the pattern from Section~\ref{sec:IMP}
\[\cfg{\tt x\!:=\!x\!+\!1 \ || \ x\!:=\!x\!+\!1}{{\tt x} \!\mapsto\!
n} \andx (n = 0 \orx n = 1)\]
is matched by the configurations with code ``${\tt x\!:=\!x\!+\!1 \ || \
x\!:=\!x\!+\!1}$'' and state mapping program variable \texttt{x} into integer
$n$ with $n$ being either $0$ or $1$.
We use \textit{italic} for mathematical variables in $\Var$ and
\texttt{typewriter} for program variables (program variables are represented in
matching logic as constants of sort $\PVar$, see Section~\ref{sec:IMP}).

The ``topmost'' matching logic variant that we use in this paper is a fragment
of the general matching logic approach proposed in
\cite{rosu-2017-lmcs,rosu-2015-rta}, where terms are regarded as predicates
only at the top level, where they have the sort $\Cfg$.
The general matching logic approach in \cite{rosu-2017-lmcs,rosu-2015-rta}
allows the distinctive terms-as-predicates view as well as the mix of logical
and structural constraints at all levels, for any terms of any sorts.
As an example, the pattern above may be equivalently written as
\[\cfg{\tt x\!:=\!x\!+\!1 \ || \ x\!:=\!x\!+\!1}{{\tt x} \!\mapsto\!
n \andx (n = 0 \orx n = 1)} \]
thus localizing the logical constraints to the actual place in the configuration
where they matter.
We remind the reader that in this paper we limit ourselves to ``topmost'' matching
logic, and that we take the freedom to drop the ``topmost'' and just call it
``matching logic''.

Next, we recall how matching logic formulae can be translated into FOL
formulae, so that its satisfaction becomes FOL satisfaction in the model of
configurations, $\cal T$.
Then, we can use conventional theorem provers or proof assistants for pattern
reasoning.

\begin{defiC}[\cite{rosu-ellison-schulte-2010-amast}] \label{dfn:ml2fol}
Let $\square$ be a fresh $\Cfg$ variable.
For a pattern $\varphi$, let $\varphi^\square$ be the FOL formula formed
from $\varphi$ by replacing
basic patterns $\pi\in {\cal T}_{\!\!\Sigma,\Cfg}(\Var)$ with equalities
$\square = \pi$.
If $\rho\! :\! \Var\! \ra\! {\cal T}$ and $\gamma \!\in\! {\cal T}_{\!\!\Cfg}$
then let the valuation $\rho^\gamma : \Var \cup \{\square\} \ra {\cal T}$
be such that
$\rho^\gamma(x) = \rho(x)$ for $x\in\Var$ and $\rho^\gamma(\square)=\gamma$.
\end{defiC}

\noindent
With the notation in Definition~\ref{dfn:ml2fol},
$\mlmodels{\gamma}{\rho}{\varphi}$ iff $\rho^\gamma \models
\varphi^\square$, and $\models \varphi$ iff ${\cal T} \models
\varphi^\square$.
Thus, matching logic is a methodological fragment of the FOL theory of
${\cal T}$.
We drop $\square$ from $\varphi^\square$ when it is
clear in context that we mean the FOL formula instead of the matching
logic pattern.
It is often technically convenient to eliminate $\square$ from $\varphi$, by
replacing $\square$ with a $\Cfg$ variable $\cvar$ and using
$\squaresubst{\varphi}{\cvar}$ instead of $\varphi$. 
We use the FOL representation in the \Step proof rule in
Fig.~\ref{fig:proof-system}, and to establish relative completeness in
Section~\ref{sec:completeness}.

\section{Specifying Reachability}\label{sec:reachability-rules}

In this section we define one-path and all-path reachability.
We begin by recalling some matching logic
reachability~\cite{rosu-stefanescu-2012-icalp} notions that we need for
specifying reachability.

\begin{defiC}[\cite{rosu-stefanescu-2012-icalp}] \label{dfn:mlr}
A (one-path) \textbf{reachability rule} is a pair $\varphi \Ra \varphi'$, where
$\varphi$ and $\varphi'$ are patterns (which can have free variables).
Rule $\varphi \Ra \varphi'$ is \textbf{weakly well-defined} iff for any
$\gamma \in \TCfg$ and $\rho : \Var \ra {\cal T}$ with
$\mlmodels{\gamma}{\rho}{\varphi}$, there exists $\gamma' \in \TCfg$ with
$\mlmodels{\gamma'}{\rho}{\varphi'}$.
A \textbf{reachability system} is a set of reachability rules.
Reachability system $\Sem$ is \textbf{weakly well-defined} iff each rule is
weakly well-defined.
$\Sem$ induces a \textbf{transition system} $({\cal T}, \RRs)$ on the
configuration model: $\gamma \RRs \gamma'$ for $\gamma, \gamma' \in \TCfg$ iff
there is some rule $\varphi \Ra \varphi'$ in $\Sem$ and some valuation $\rho :
\Var \ra {\cal T}$ with $\mlmodels{\gamma}{\rho}{\varphi}$ and
$\mlmodels{\gamma'}{\rho}{\varphi'}$.
A \textbf{$\RRs$-path} is a finite sequence $\gamma_0 \!\RRs\! \gamma_1 \!\RRs\!
\ldots \!\RRs\! \gamma_n$ with $\gamma_0,\! \ldots,\! \gamma_n \in \TCfg$.
A $\RRs$-path is \textbf{complete} iff it is not a strict prefix of any other
$\RRs$-path.
\end{defiC}

We assume an operational semantics is a set of (unconditional) reduction rules
``$l \Ra r \ {\sf if} \ b$'', where $l,r \in T_{\Sigma, \Cfg}(\Var)$ are program
configurations with variables and $b \in T_{\Sigma, \Bool}(\Var)$ is a condition
constraining the variables of $l, r$.
Styles of operational semantics using only such (unconditional) rules include
evaluation contexts~\cite{DBLP:books/daglib/0023092}, the chemical abstract
machine~\cite{DBLP:journals/tcs/BerryB92} and \K~\cite{rosu-serbanuta-2010-jlap}
(see Section~\ref{sec:IMP} for an evaluation contexts semantics).
Several large languages have been given semantics in such styles, including
C~\cite{ellison-rosu-2012-popl} (more than 3000 rules) and R5RS
Scheme~\cite{scheme-redex}.
The reachability proof system works with any set of rules of this form,
being agnostic to the particular style of semantics.

Such a rule ``$l \Ra r \ {\sf if} \ b$'' states that a ground configuration
$\gamma$ which is an instance of $l$ and satisfies the condition $b$ reduces to
an instance $\gamma'$ of $r$.
Matching logic can express terms with constraints: $l \andx b$ is
satisfied by exactly the $\gamma$ above.
Thus, we can regard such a semantics as a particular weakly well-defined
reachability system ${\cal S}$ with rules of the form ``$l \andx b \Ra r$''.
The weakly well-defined condition on $\Sem$ guarantees that if $\gamma$ matches
the left-hand-side of a rule in $\Sem$, then the respective rule induces
an outgoing transition from $\gamma$.
The transition system induced by $\Sem$ describes precisely the
behavior of any program in any given state.
In~\cite{rosu-stefanescu-2012-icalp,rosu-stefanescu-2012-fm,rosu-stefanescu-2012-oopsla}
we show that reachability rules capture one-path reachability properties and
Hoare triples for deterministic languages.

Formally, let us fix an operational semantics given as a reachability system
$\Sem$.
Then, we can specify reachability in the transition system induced by $\Sem$,
both all-path and one-path, as follows:

\begin{mydefinition}\label{def:partial-correctness}
An \textbf{all-path reachability rule} is a pair $\varphi \ral \varphi'$ of
patterns $\varphi$ and $\varphi'$.

An all-path reachability rule $\varphi \ral \varphi'$ is satisfied,
\mbox{${\cal S} \models \varphi \ral \varphi'$},
iff for all
complete $\RRs$-paths $\tau$ starting with $\gamma \in \TCfg$ and for
all $\rho : \Var \ra {\cal T}$ such that $\mlmodels{\gamma}{\rho}{\varphi}$,
there exists some $\gamma' \in \tau$ such that
$\mlmodels{\gamma'}{\rho}{\varphi'}$.

A one-path reachability rule $\varphi \rex \varphi'$ is satisfied,
\mbox{${\cal S} \models \varphi \rex \varphi'$}, iff for all
$\gamma \in \TCfg$ and $\rho : \Var \ra {\cal T}$ such that
$\mlmodels{\gamma}{\rho}{\varphi}$, there is either a $\RRs$-path from
$\gamma$ to some $\gamma'$ such that $\mlmodels{\gamma'}{\rho}{\varphi'}$,
or there is a diverging execution
$\gamma \RRs \gamma_1 \RRs \gamma_2 \RRs \cdots$ from $\gamma$.
\end{mydefinition}

The racing increment example in Section~\ref{sec:IMP} can be specified by
\[
\cfg{\tt x\!:=\!x\!+\!1 \ || \ x\!:=\!x\!+\!1}{{\tt x} \!\mapsto\! m}
\ral \exists n \ (\cfg{\pvar{skip}}{{\tt x} \!\mapsto\! n}
\andx (n = m +_\Int 1 \orx n = m +_\Int 2)
\]
which states that every terminating execution reaches a state where
execution of both threads is complete and the value of ${\tt x}$ has
increased by $1$ or $2$ (this code has a race).

A Hoare triple describes the resulting state after execution finishes,
so it corresponds to a reachability rule where the right side contains
no remaining code.
However, all-path reachability rules are strictly more expressive than Hoare
triples, as they can also specify intermediate configurations (the code in the
right-hand-side need not be empty).
%
Reachability rules provide a unified representation for both language semantics
and program specifications: $\varphi \rex \varphi'$ for semantics
and $\varphi \ral \varphi'$ for all-path reachability specifications.
%
Note that, like Hoare triples, reachability rules can only specify properties of
complete paths (that is, terminating execution paths).
One can use existing Hoare logic techniques to break reasoning about
a non-terminating program into reasoning about its terminating components.

\section{Reachability Proof System}\label{sec:proof-system}

\begin{figure}[t]
\begin{small}
\PROOFSYSTEM
\end{small}
\caption{\label{fig:proof-system} Proof system for reachability.
We make the standard assumption that the free variables of $\lhs \Ra \rhs$ in
the \Step proof rule are fresh (e.g., disjoint from those of $\varphi \Rra
\varphi'$).}
\end{figure}

Fig.~\ref{fig:proof-system} shows our novel proof system for all-path
reachability.
The target language is given as a weakly well-defined reachability system $\Sem$.
The soundness result (Thm.~\ref{thm:soundness}) guarantees that $\Sem \models
\varphi \Rra \varphi'$ if $\ssequent{\varphi}{\varphi'}$ is derivable.
Note that the proof system derives more general sequents of the form
$\sequent[\cal C]{\cal A}{\varphi}{\varphi'}$, where ${\cal A}$ and ${\cal C}$
are sets of reachability rules.
The rules in ${\cal A}$ are called {\em axioms} and rules in ${\cal C}$ are
called {\em circularities}.
If either ${\cal A}$ or ${\cal C}$ does not appear in a sequent, it means the
respective set is empty: $\ssequent[\cal C]{\varphi}{\varphi'}$ is a shorthand
for $\sequent[\cal C]{\emptyset}{\varphi}{\varphi'}$, and $\sequent{\cal
A}{\varphi}{\varphi'}$ is a shorthand for $\sequent[\emptyset]{\cal
A}{\varphi}{\varphi'}$, and
$\ssequent{\varphi}{\varphi'}$ is a shorthand for
$\sequent[\emptyset]{\emptyset}{\varphi}{\varphi'}$.
Initially, both ${\cal A}$ and ${\cal C}$ are empty.
Note that ``$\rightarrow$'' in \Step and \Consequence denotes logical implication.

The intuition is that the reachability rules in ${\cal A}$ can be assumed valid,
while those in ${\cal C}$ have been postulated but not yet justified.
After making progress from $\varphi$ (at least one derivation by \Step or by
\Axiom with the rules in ${\cal A}$), the rules in ${\cal C}$ become
(coinductively) valid (can be used in derivations by \Axiom).
During the proof, circularities can be added to ${\cal C}$ via \Circularity,
flushed into ${\cal A}$ by \Transitivity, and used via \Axiom.
The desired semantics for sequent $\sequent[\cal C]{\cal A}{\varphi}{\varphi'}$
(read ``$\Sem$ with axioms ${\cal A}$ and circularities $\cal C$ proves $\varphi
\Rra \varphi'$'') is: $\varphi \Rra \varphi'$ holds if the rules in ${\cal
A}$ hold and those in ${\cal C}$ hold after taking at least one step from
$\varphi$ in the transition system $(\RRs, {\cal T})$, and if ${\cal
C}\neq\emptyset$ then $\varphi$ reaches $\varphi'$ after at least one step on
all complete paths.
As a consequence of this definition, any rule $\varphi \Rra \varphi'$ derived by \Circularity has
the property that $\varphi$ reaches $\varphi'$ after at lest one step, due to
\Circularity having a prerequisite $\sequent[{\cal C} \cup \{ \varphi \Rra
\varphi' \}]{\cal A}{\varphi}{\varphi'}$ (with a non-empty set of
circularities).
We next discuss the proof rules.

\Step derives a sequent where $\varphi$ reaches $\varphi'$ in one step on all
paths.
The first premise ensures any configuration matching $\varphi$ matches the
left-hand-side $\lhs$ of some rule in $\Sem$ and thus, as $\Sem$ is weakly
well-defined, can take a step.
Formally, if $\mlmodels{\gamma}{\rho}{\varphi}$, then there exists some rule
$\lhs \Ra \rhs \in \Sem$ and some valuation $\rho'$ of the free variables of
$\lhs$ such that $\mlmodels{\gamma}{\rho'}{\lhs}$, and thus $\gamma$ has at
least one $\RRs$-successor generated by the rule $\lhs \Ra \rhs$.
%
The second premise ensures that each $\RRs$-successor of a configuration
matching $\varphi$ matches $\varphi'$.
Formally, if $\gamma \RRs \gamma'$ and $\gamma$ matches $\varphi$ then there is
some rule $\lhs \Ra \rhs \in \Sem$ and $\rho : \Var \ra {\cal T}$ such that
$\mlmodels{\gamma}{\rho}{\varphi \land \lhs}$ and
$\mlmodels{\gamma'}{\rho}{\rhs}$; then the second premise implies $\gamma'$ matches
$\varphi'$.

Designing a proof rule for deriving an execution step along all paths is
non-trivial.
For instance, one might expect \Step to require as many premises as there are
transitions going out of $\varphi$, as is the case for the examples presented
later in this paper.
However, that is not possible, as the number of successors of a configuration
matching $\varphi$ may be unbounded even if each matching configuration has
a finite branching factor in the transition system.
\Step avoids this issue by requiring only one premise for each rule by which some
configuration $\varphi$ can take a step, even if that rule can be used to derive
multiple transitions.
To illustrate this situation, consider a language defined by $\Sem \equiv \{
\cfgone{n_1} \andx n_1 >_\Int n_2 \Ra \cfgone{n_2} \}$, with $n_1$ and $n_2$
non-negative integer variables.
A configuration in this language is a singleton with a non-negative integer.
Intuitively, a positive integer transits into a strictly smaller non-negative
integer, in a non-deterministic way.
The branching factor of a non-negative integer is its value.
Then $\Sem \models \cfgone{m} \Rra \cfgone{0}$.
Deriving it
reduces (by \Circularity and other proof rules) to deriving
$\cfgone{m_1} \andx m_1 >_\Int 0 \Rra \exists m_2 \ (\cfgone{m_2} \andx m_1
>_\Int m_2)$.
The left-hand-side is matched by any positive integer, and thus its branching
factor is infinity.
Deriving this rule with \Step requires only two premises,
$\models (\cfgone{m_1} \andx m_1
>_\Int 0) \ra \exists n_1 n_2 \ ( \cfgone{n_1} \andx n_1 >_\Int n_2)$ and
$\models \exists \cvar \ (\cvar = \cfgone{m_1} \andx m_1 >_\Int 0 \andx
\cvar = \cfgone{n_1} \andx n_1 >_\Int n_2) \andx \cfgone{n_2} \ra \exists m_2 \
(\cfgone{m_2} \andx m_1 >_\Int m_2)$.
A similar situation arises in real life for
languages with thread pools of arbitrary size.

\Axiom applies a trusted rule.
\Reflexivity and \Transitivity capture the corresponding closure properties of the
reachability relation.
\Reflexivity requires ${\cal C}$ to be empty to ensure that all-path rules
derived with non-empty ${\cal C}$ take at least one step.
\Transitivity enables the circularities as axioms for the second premise, since
if ${\cal C}$ is not empty, the first premise is guaranteed to take at least one
step.
\Consequence, \Case and \Abstraction are adapted from Hoare logic.
Ignoring circularities, these seven rules discussed so far constitute formal
infrastructure for symbolic execution.

%

\Circularity has a coinductive nature, allowing us to make new circularity
claims.
We typically make such claims for code with repetitive behaviors, such as loops,
recursive functions, jumps, etc.
If there is a derivation of the claim using itself as a circularity, then the
claim holds.
This would obviously be unsound if the new assumption was available immediately,
but requiring progress (taking at least one step in the transition system $({\cal
T}, \RRs)$) before circularities can be used ensures that only diverging
executions can correspond to endless invocation of a circularity.

One important aspect of concurrent program verification, which we do not
address in this paper, is proof compositionality.
Our focus here is limited to establishing a sound and complete
language-independent proof system for all-path reachability rules, to serve as a
foundation for further results and applications, and to discuss our current
implementation of it.
We only mention that we have already studied proof compositionality for earlier
one-path variants of reachability logic \cite{rosu-stefanescu-2012-fm}, showing
that there is a mechanical way to translate any Hoare logic proof derivation
into a reachability proof of similar size and structure, but based entirely on the
operational semantics of the language.
The overall conclusion of our previous study, which we believe will carry over to
all-path reachability, was that compositional reasoning can be achieved methodologically
using our proof system, by proving and then using appropriate reachability rules
as lemmas.
However, note that this works only for theoretically well-behaved languages which
enjoy a compositional semantics.
For example, a language whose semantics assumes a bounded heap size, or
which has constructs whose semantics involve the entire program, e.g., call/cc,
will lack compositionality.


\section{Verifying Programs}\label{sec:IMP}

\begin{figure}[t]
\begin{small}
\input{imp-semantics}
\end{small}
\caption{\label{fig:IMP}\IMP language syntax and operational semantics based on
evaluation contexts.}
\end{figure}

In this section we show a few examples of using our proof system to verify
programs based on an operational semantics.
In a nutshell, the proof system enables generic symbolic execution combined with
circular reasoning.
Symbolic execution is achieved by rewriting modulo domain reasoning.

First, we introduce a simple parallel imperative language, \IMP.
Fig.~\ref{fig:IMP} shows its syntax and an operational semantics based on
evaluation contexts~\cite{DBLP:books/daglib/0023092} (we choose evaluation
contexts for presentation purposes only). 
\IMP has only integer expressions.
When used as conditions of {\tt if} and {\tt while}, zero means false and any
non-zero integer means true (like in C).
Expressions are formed with integer constants, program variables, and
conventional arithmetic constructs.
Arithmetic operations are generically described as {\tt op}.
\IMP statements are assignment, {\tt if}, {\tt while}, sequential composition
and parallel composition.
\IMP has shared memory parallelism without explicit synchronization.
The examples use the parallel construct only at the top-level of the programs.
The second example shows how to achieve synchronization using the existing
language constructs.

The program configurations of \IMP are pairs $\cfg{\tt code}{\sigma}$, where
{\tt code} is a program fragment and $\sigma$ is a state term mapping program
variables into integers.
As usual, we assume appropriate definitions for the integer and map domains
available, together with associated operations like arithmetic operations ($i_1
\, {\it op}_{\it Int} \, i_2$, etc.) on the integers and lookup ($\sigma(x)$) and
update ($\sigma[x \leftarrow i]$) on the maps.
We also assume a context domain with a plugging operation ($C[t]$) that composes
a context and term back into a configuration.
A configuration context consists of a code context and a state.
The definition in Fig.~\ref{fig:IMP} consists of eight reduction rules
between program configurations, which make use of first-order variables: $x$ is
a variable of sort $\PVar$; $e$ is a variable of sort $\Exp$; $s,s_1,s_2$ are
variables of sort $\Stmt$; $i,i_1,i_2$ are variables of sort $\Int$; $\sigma$ is
a variable of sort $\State$; $C$ is a variable of sort $\Context$.
A rule reduces a configuration by splitting it into a context and a redex,
rewriting the redex and possibly the context, and then plugging the resulting
term into the resulting context.
As an abbreviation, a context is not mentioned if not used; e.g., the rule {\bf
op} is in full $\cfg{C}{\sigma}[i_1 \, {\tt op} \, i_2] \Ra
\cfg{C}{\sigma}[i_1 \, {\it op}_{\it Int} \, i_2]$.
For example, configuration $\cfg{{\tt x :=} \, (2 + 5) - 4}{\sigma}$
reduces to $\cfg{{\tt x :=} \, 7 - 4}{\sigma}$ by applying the ${\bf op}_+$
rule with $C \equiv {\tt x :=} \, \blacksquare - 4$, $\sigma \equiv \sigma$,
$i_1 \equiv 2$ and $i_2 \equiv 5$.
We can regard the operational semantics of \IMP above as a set of reduction
rules of the form ``$l \Ra r \ {\sf if} \ b$'', where $l$ and $r$ are program
configurations with variables constrained by boolean condition $b$.
%
%
As discussed in Section~\ref{sec:reachability-rules}, our proof system works
with any rules of this form.

\newif\ifcompactstates
\compactstatesfalse

\ifcompactstates
\newcommand{\startpgm}{\pvar{x}+}
\newcommand{\addpgm}[1]{#1+}
\newcommand{\storepgm}[1]{#1+}
\newcommand{\donepgm}{\pvar{skip}}
\newcommand{\ecfg}[3]{\varphi^{#3}_{#1 \pars #2}}
\else
\newcommand{\startpgm}{\pvar{x} := \pvar{x} + 1}
\newcommand{\addpgm}[1]{\pvar{x} := #1 + 1}
\newcommand{\storepgm}[1]{\pvar{x} := #1}
\newcommand{\donepgm}{\pvar{skip}}
\newcommand{\ecfg}[3]{
{\left\langle
{}^{#1 \pars #2,}_{\pvar{x} \mapsto #3}
\right\rangle}}
\fi

\newcommand{\pars}{\mathop{\mathbf{||}}}

\newcommand{\startpat}{\ecfg{\addpgm{m}}{\startpgm}{m}}
\newcommand{\lgoalpat}{\ecfg{\storepgm{m+_\Int 1}}{\startpgm}{m}}
\newcommand{\rgoalpat}{\ecfg{\addpgm{m}}{\addpgm{m}}{m}}

\begin{figure}[t]
\ifcompactstates
\begin{minipage}[c]{0.6\linewidth}
\fi
\fontsize{.8em}{.8em}\selectfont 
\begin{tikzcd}[column sep=tiny,row sep=tiny]
\ecfg{\startpgm}{\startpgm}{m} \arrow[start anchor={[yshift=1ex]}]{d}\arrow{r}
  & \ecfg{\startpgm}{\addpgm{m}}{m} \arrow[start anchor={[yshift=1ex]}]{d}\arrow{r}
    & \ecfg{\startpgm}{\storepgm{m+_\Int 1}}{m} \arrow[start anchor={[yshift=1ex]}]{d}\arrow{r}
      & \ecfg{\startpgm}{\donepgm}{m+_\Int 1} \arrow[bend left]{dr}
        & \\[-1ex]
\ecfg{\addpgm{m}}{\startpgm}{m} \arrow[start anchor={[yshift=1ex]}]{d}\arrow{r}
  & \ecfg{\addpgm{m}}{\addpgm{m}}{m}\arrow[start anchor={[yshift=1ex]}]{d}\arrow{r}
    & \ecfg{\addpgm{m}}{\storepgm{m+_\Int 1}}{m}\arrow[start anchor={[yshift=1ex]}]{d}\arrow{r}
      & \ecfg{\addpgm{m}}{\donepgm}{m+_\Int 1}\arrow[start anchor={[yshift=.5ex]},end anchor={[yshift=-.5ex]}]{d}
        & \ecfg{\addpgm{m+_\Int 1}}{\donepgm}{m+_\Int 1}\arrow[start anchor={[yshift=.5ex]},end anchor={[yshift=-.5ex]}]{d} \\[-1ex]
\ecfg{\storepgm{m+_\Int 1}}{\startpgm}{m} \arrow[start anchor={[yshift=1ex]}]{d}\arrow{r}
  & \ecfg{\storepgm{m+_\Int 1}}{\addpgm{m}}{m}\arrow[start anchor={[yshift=1ex]}]{d}\arrow{r}
    & \ecfg{\storepgm{m+_\Int 1}}{\storepgm{m+_\Int 1}}{m}\arrow[start anchor={[yshift=1ex]}]{d}\arrow{r}
      & \ecfg{\storepgm{m+_\Int 1}}{\donepgm}{m+_\Int 1}\arrow[start anchor={[yshift=.5ex]},end anchor={[yshift=-.5ex]}]{d}
        & \ecfg{\storepgm{m+_\Int 2}}{\donepgm}{m+_\Int 1}\arrow{dd} \\[-1ex]
\ecfg{\donepgm}{\startpgm}{m+_\Int 1}\arrow[bend right,end anchor={[yshift=2ex]}]{dr}
  & \ecfg{\donepgm}{\addpgm{m}}{m+_\Int 1} \arrow{r}
    & \ecfg{\donepgm}{\storepgm{m+_\Int 1}}{m+_\Int 1} \arrow{r}
      & \ecfg{\donepgm}{\donepgm}{m+_\Int 1} 
         & \\[-1ex]
  & \ecfg{\donepgm}{\addpgm{m+_\Int 1}}{m+_\Int 1}\arrow{r}
    & \ecfg{\donepgm}{\storepgm{m+_\Int 2}}{m+_\Int 1}\arrow{rr}
      & 
        & \ecfg{\donepgm}{\donepgm}{m+_\Int 2} \\
\end{tikzcd}

\ifcompactstates
\end{minipage}
\hspace{0.5cm}
\begin{minipage}[c]{0.3\linewidth}
\begin{align*}
\ecfg{p_1}{p_2}{n} & = \cfg{p_1 \pars p_2}{{\tt x} \mapsto n} \\
\startpgm & = {\tt x} := {\tt x} + 1 \\
\addpgm{i} & = {\tt x} := i + 1 \\
\storepgm{i}  & = {\tt x} := i \\
\donepgm & = \mbox{\tt skip} \\
\end{align*}
\end{minipage}
\fi
\caption{State space of the racing increment example}\label{fig:increment-states}
\end{figure}


Next, we illustrate the proof system on a few examples.
The first example shows that our proof system enables exhaustive state
exploration, similar to symbolic model-checking but based on the operational
semantics.
Although humans prefer to avoid such explicit proofs and instead
methodologically use abstraction or compositional reasoning whenever possible
(and such methodologies are not excluded by our proof system), a complete proof
system must nevertheless support them.
%
The code ${\tt x}\!:=\!{\tt x}\!+\!1 \ || \ {\tt x}\!:=\!{\tt x}\!+\!1$ exhibits
a race on ${\tt x}$: the value of ${\tt x}$ increases by $1$ when both reads
happen before either write, and by $2$ otherwise.
The all-path rule that captures this behavior is
\[\cfg{{\tt x}\!:=\!{\tt x}\!+\!1 \ || \ {\tt x}\!:=\!{\tt x}\!+\!1}{{\tt x}
\!\mapsto\! m} \Rra \exists n \ (\cfg{\tt skip}{{\tt x} \!\mapsto\! n}
\andx (n = m +_\Int 1 \orx n = m +_\Int 2)\]
We show that the program has exactly these behaviors by
deriving this rule in the proof system.
Call the right-hand-side pattern $G$.
The proof contains subproofs of $c \Rra G$ for every reachable
configuration $c$, tabulated in Fig.~\ref{fig:increment-states}.
%
The subproofs for $c$ matching $G$ use \Reflexivity and \Consequence, while the
rest use \Transitivity, \Step, and \Case to reduce to the proofs for the next
configurations.
For example, the proof fragment below shows how $\cfg{{\tt x} := m + 1 \pars
{\tt x} := {\tt x} + 1}{{\tt x} \mapsto m} \Rra G$ reduces to $\cfg{{\tt x} := m
+_\Int 1 \pars {\tt x} := {\tt x} \!+\! 1}{{\tt x} \mapsto m} \Rra G$ and
$\cfg{{\tt x} := m \!+\! 1 \pars {\tt x} := m \!+\! 1}{{\tt x} \mapsto m} \Rra
G$:
\[
  {\small
\fontsize{.7em}{.7em}\selectfont 
\inferrule*[right=\textsc{Trans}, sep=5ex]
 {{\inferrule*[left=\Step]{\ldots}
  {\mbox{$\begin{array}{rl}\startpat \Rra & \lgoalpat \\
                                     \lor & \rgoalpat\end{array}$}}
  \hspace{1em}
  \inferrule*[Right=\textsc{CA}, sep=3ex]
  {\inferrule*{\ldots}{\lgoalpat \!\!\Rra\! G}
   \inferrule*{\ldots}{\rgoalpat \!\!\Rra\! G}}
  {\lgoalpat\!\lor\!\rgoalpat \Rra G}}
  }
 {\cfg{{\tt x} := m + 1 \pars {\tt x} := {\tt x} + 1}{{\tt x} \mapsto
     m} \Rra G}
}\]
For the rule hypotheses of \Step{} above, note that all rules but
\textbf{lookup} and $\mbox{\textbf{op}}_+$ make the overlap condition $\exists c
\ \left(\ecfg{\addpgm{m}}{\startpgm}{m}[c / \square] \land \varphi_l[c /
\square]\right)$ unsatisfiable, and only one choice of free variables works for
the \textbf{lookup} and $\mbox{\textbf{op}}_+$ rules.
For \textbf{lookup}, $\varphi_l$ is $\langle C, \sigma \rangle[x]$ and the
overlap condition is only satisfiable if the logical variables $C$, $\sigma$
and $x$ are equal to $({\tt x} := m + 1 \pars {\tt x} := \blacksquare + 1)$,
$({\tt x} \mapsto m)$, and ${\tt x}$, respectively.
Under this assignment, the pattern $\varphi_r = \langle C,\sigma\rangle[\sigma(x)]$
is 
equivalent to $\cfg{{\tt x} := m + 1 \pars {\tt x} := m + 1}{{\tt x} \mapsto m}$,
the right branch of the disjunction.
The $\mbox{\textbf{op}}_+$ rule is handled similarly.
The assignment for \textbf{lookup} can also witness the existential in
the progress hypothesis of \Step{}.
Subproofs for other states in Fig.~\ref{fig:increment-states} can be
constructed similarly.

\begin{wrapfigure}{|L}{6.25cm}
\lstset{mathescape}
\begin{minipage}{2.75cm}
\scriptsize
\begin{lstlisting}
f0 = 1;
turn = 1;
while (f1 && turn)
  skip
x = x + 1;
f0 = 0;
\end{lstlisting}
\end{minipage}
\begin{minipage}{3cm}
\scriptsize
\begin{lstlisting}
f1 = 1;
turn = 0;
while (f0 && (1 - turn))
  skip
x = x + 1;
f1 = 0;
\end{lstlisting}
\end{minipage}
\caption{\label{fig:dekker}Peterson's algorithm (threads ${\tt T_0}$ and ${\tt
T_1}$)}
\end{wrapfigure}

The next two examples use loops and thus need to state and prove invariants.
As discussed in~\cite{rosu-stefanescu-2012-oopsla}, \Circularity generalizes the
various language-specific invariant proof rules encountered in Hoare logics.
One application is reducing a proof of $\varphi \Rra \varphi'$ to proving
$\varphi_{\rm inv} \Rra \varphi_{\rm inv} \lor \varphi'$ for some pattern
invariant $\varphi_{\rm inv}$.
We first show $\models \varphi \ra \varphi_{\rm inv}$, and use \Consequence to
change the goal to $\varphi_{\rm inv} \Rra \varphi'$.
This is claimed as a circularity, and then proved by transitivity with
$\varphi_{\rm inv} \lor \varphi'$.
The second hypothesis $\{ \varphi_{\rm inv} \Rra \varphi' \} \vdash \varphi_{\rm
inv} \lor \varphi' \Rra \varphi'$ is proved by \Case, \Axiom, and \Reflexivity.

Next, we can use Peterson's algorithm for mutual exclusion to eliminate the race
as shown in Fig.~\ref{fig:dekker}.
The all-path rule $\varphi \Rra \varphi'$ that captures the new behavior is
\[\begin{array}{rl}
& \cfg{{\tt T_0} \pars {\tt T_1}} {(\pvar{f0} \mapsto 0,\ \pvar{f1} \mapsto 0,\
\pvar{x} \mapsto N)} \\
\Rra & \exists t \
\cfg{\texttt{skip}}{(\pvar{f0} \mapsto 0,\ \pvar{f1} \mapsto 0,\ \pvar{x}
\mapsto N +_\Int 2,\ \pvar{turn} \mapsto t)} \\
\end{array}\]
Similarly to the unsynchronized example, the proof contains subproofs of $c \Rra
\varphi'$ for every configuration $c$ reachable from $\varphi$.
The main difference is that \Circularity is used with each of these rules $c
\Rra \varphi'$ with one of the two threads of $c$ in the {\tt while} loop (these
rules capture the invariants).
Thus, when we reach a configuration $c$ visited before, we use the rule added by
\Circularity to complete the proof. 

The final example is the program \texttt{SUM} $\equiv$ ``\texttt{s := 0; LOOP}''
(where $\texttt{LOOP}$ stands for ``\texttt{while (n>0) (s := s+n; n :=
n-1)}''), which computes in $\texttt{s}$ the sum of the numbers from $1$ up to
$\texttt{n}$.
The all-path reachability rule $\varphi \Rra\! \varphi'$ capturing this behavior
is 
\[\cfg{\texttt{SUM}}{(\texttt{s} \mapsto s,\ \texttt{n} \mapsto n)} \andx n
\geq_\Int 0 \ \Rra \ \cfg{{\tt skip}}{(\texttt{s} \mapsto n *_\Int (n +_\Int 1)
/_\Int 2,\ \texttt{n} \mapsto 0)}\]
We derive the above rule in our proof system by using \Circularity with the
invariant rule
$$\exists n' (\cfg{\texttt{LOOP}}{(\texttt{s} \!\mapsto\! (n
\!-_\Int\! n') \!*_\Int\! (n \!+_\Int\! n' \!+_\Int\! 1) /_\Int 2,\ \texttt{n}
\!\mapsto\! n')} \andx n' \!\geq_\Int\! 0) \Rra \varphi'$$
Previous
work~\cite{rosu-stefanescu-2012-icalp,rosu-stefanescu-2012-fm,rosu-stefanescu-2012-oopsla,rosu-stefanescu-ciobaca-moore-2013-lics}
presented a proof system able to derive similar rules, but which hold along {\it
some} execution path, requiring a separate proof that the program is
deterministic.

\section{Implementation}\label{sec:implementation}

Here we briefly discuss our prototype implementation of the proof system in
Fig.~\ref{fig:proof-system} in \K~\cite{rosu-serbanuta-2010-jlap}.
We choose \K because it is a modular semantic language design framework, it is
used for teaching programming languages at several universities, and there are
several languages defined in it including C~\cite{ellison-rosu-2012-popl},
PHP~\cite{filaretti-maffeis-2014-ecoop}, Python, and Java.
For brevity, we do not present \K here.
We refer the reader to~\url{http://kframework.org} for language definitions, a
tutorial, and our prototype.
As discussed in Section~\ref{sec:reachability-rules}, we simply view a \K
semantics as a set of reachability rules of the form ``$l \andx b \Ra r$''.

The prototype is implemented in Java, and uses
Z3~\cite{DBLP:conf/tacas/MouraB08} for domain reasoning.
It takes an operational semantics and uses it to perform concrete or symbolic
execution.
At its core, it performs {\em narrowing} of a conjunctive pattern with
reachability rules between conjunctive patterns, where a conjunctive pattern is
a pattern of the form $\exists X (\pi \andx \psi)$, with $X$ a set of variables,
$\pi$ a basic pattern (program configurations with variables), and $\psi$ a
structureless formula.
Narrowing is necessary when a conjunctive pattern is too abstract to match the
left-hand side of any rule, but is unifiable with the left-hand sides of some
rules.
For instance, consider the \IMP code fragment ``$\texttt{if (} b \texttt{) then
x = 1; else x = 0;}$''.
This code does not match the left-hand sides of either of the two rules giving
semantics to {\tt if} (similar to $\textbf{cond}_1$ and $\textbf{cond}_2$ in
Fig.~\ref{fig:IMP}), but it is unifiable with the left-hand sides of both rules.
Intuitively, if we use the rules of the semantics, taking steps of rewriting on
a ground configuration yields concrete execution, while taking steps of
narrowing on a conjunctive pattern yields symbolic execution.
In our practical evaluation, we found that conjunctive patterns tend to suffice
to specify both the rules for operational semantics and program specifications.

For each step of narrowing, the \K engine uses {\em unification modulo
theories}.
In our implementation, we distinguish a number of mathematical theories (e.g.
booleans, integers, sequences, sets, maps, etc) which the underlying SMT solver
can reason about.
Specifically, when unifying a conjunctive pattern $\exists X (\pi
\andx \psi)$ with the left-hand side of a rule $\exists X_l (\pi_l \andx
\psi_l)$ (we assume $X \cap X_l = \emptyset$), the \K engine begins with the
syntactic unification of the basic patterns $\pi$ and $\pi_l$.
Upon encountering corresponding subterms ($\pi'$ in $\pi$ and $\pi_l'$ in
$\pi_l$) which are both terms of one of the theories above, it records an equality
$\pi' = \pi_l'$ rather than decomposing the subterms further (if one is in a
theory, and the other one is in a different theory or is not in any theory, the
unification fails).
If this stage of unification is successful, we end up with a conjunction
$\psi_u$ of constraints, some having a variable in one side and some with both sides
in one of the theories.
Satisfiability of $\exists X \cup X_l (\psi \andx \psi_u \andx \psi_l)$ is
then checked by the SMT solver.
If it is satisfiable, then narrowing takes a step from $\exists X (\pi \andx
\psi)$ to $\exists X \cup X_l \cup X_r (\pi_r \andx \psi \andx \psi_u \andx
\psi_l \andx \psi_r)$, where $\exists X_r (\pi_r \andx \psi_r)$ is the
right-hand side of the rule.
Intuitively, ``collecting'' the constraints $\psi_u \andx \psi_l \andx \psi_r$
is similar to collecting the path constraint in traditional symbolic execution
(but is done in a language-generic manner).
For instance, in the {\tt if} case above, narrowing with the two semantics
rules results in collecting the constraints $b = \texttt{true}$ and $b =
\texttt{false}$.

The \K engine accepts a set of user provided rules to prove together, which
capture the behavior of the code being verified.
Typically, these rules specify the behavior of recursive functions and while
loops.
For each rule, the \K engine searches starting from the left-hand side for
formulae which imply the right-hand side, starting with ${\cal S}$ the
semantics and ${\cal C}$ all the rules it attempts to prove.
By a derived proof rule called \emph{Set Circularity}, this suffices to show that
each rule is valid.
As an optimization, \Axiom is given priority over \Step (use specifications
rather than stepping into the code).

Most work goes into implementing the \Step proof rule, and in particular
calculating how $\rho \models \exists c \ (\varphi[c/\square] \andx
\lhs[c/\square])$ can be satisfied.
This holds when $\rho^\gamma \models \varphi$ and $\rho^\gamma \models \lhs$,
which can be checked with unification modulo theories.
To use \Step in an automated way, the \K tool constructs
 $\varphi'$ for a given $\varphi$ as a disjunction of $\rhs \andx \psi_u
\andx \psi \andx \psi_l$ over each rule $\lhs \Ra \rhs \in {\cal S}$ and each
way $\psi_u$ of unifying $\varphi$ with $\lhs$.
As discussed in Section~\ref{sec:proof-system}, in general this disjunction may
not be finite, but it is sufficient for the examples that we considered.
The \Consequence proof rule also requires unification modulo theories, to check
validity of the implication hypothesis $\models \varphi_1 \ra \varphi_1'$.  The
main difference from \Step is that the free variables of $\varphi'$ become
universality quantified when sending the query to the SMT solver.  The
implementation of the other proof rules is straight-forward.

\section{Soundness}\label{sec:soundness}

Here we discuss the soundness of our proof system.
Unlike the similar results for Hoare logics and dynamic logics, which are
separately proved for each language taking into account the particularities of
that language, we prove soundness once and for all languages.

Soundness states that a syntactically derivable sequent holds semantically.
Because of the utmost importance of the result below, we have also mechanized
its proof.
Our complete Coq formalization can be found
at~\url{http://fsl.cs.illinois.edu/rl}. 

\begin{mytheorem}{Soundness}\label{thm:soundness}
If ${\cal S} \vdash \varphi \Rra \varphi'$ then ${\cal S} \models \varphi \Rra
\varphi'$.
\end{mytheorem}
\begin{proof}
Due to \Circularity and \Transitivity the sets $\cal A$ and $\cal C$
may not remain empty, so we need to prove a more general statement
(Lemma~\ref{lem:soundness} below), including hypotheses making some
semantic assumptions about the rules in $\cal A$ and $\cal C$.

To express the different assumptions needed about the available rules
in $\cal A$ and the still-unavailable rules in $\cal C$
we define a more specific satisfaction relation.
Let $\delta \in \{+, *\}$ be a flag
and let $n \in \mathbb{N}$ be a natural number.
We define the new 
satisfaction relation ${\cal S} \models_n^\delta \varphi \ral \varphi'$
by restricting the paths in the definition of
${\cal S} \models \varphi \ral \varphi'$
to length at most $n$, and requiring that $\varphi$ be reached
after at least one step if $\delta = +$.
Formally, we define ${\cal S} \models_n^\delta \varphi \Rra \varphi'$ to hold
iff for any complete path $\tau = \gamma_1 \ldots \gamma_k$ of length
$k \leq n$ and
for any $\rho$ such that $\mlmodels{\gamma_1}{\rho}{\varphi}$, there exists $i
\in \{1, \ldots, k\}$ such that $\mlmodels{\gamma_i}{\rho}{\varphi'}$
and also that this $i$ is not 1 (i.e. $\gamma_1$ makes progress) if $\delta = +$.

Observe that
${\cal S} \models \varphi \ral \varphi'$ iff
${\cal S} \models^{*}_{n} \varphi \ral \varphi'$ for all $n \in \mathbb{N}$.
The forward implication holds because a complete path
$\tau = \gamma_1 \ldots \gamma_k$ of length $k \le n$ is
a complete path simpliciter.
The reverse implication holds because given any complete path
$\tau = \gamma_1 \ldots \gamma_k$ there are choices (such as $k$) of
$n \in \mathbb{N}$ with $k \le n$.
Theorem~\ref{thm:soundness} then 
follows from Lemma~\ref{lem:soundness}
by noting that the hypotheses about $\cal A$ and $\cal C$ are vacuously true
when those sets are empty.
\end{proof}

\begin{mylemma}\label{lem:soundness}

  If $\sequent[\cal C]{\cal A}{\varphi}{\varphi'}$ then for any $n$,
  ${\cal S} \models^+_{n} {\cal A}$ and ${\cal S} \models^+_{n - 1} {\cal C}$
  imply that ${\cal S} \models_{n}^{*} \varphi \ral \varphi'$,
  and also that ${\cal S} \models_{n}^{+} \varphi \ral \varphi'$ if
  $\cal C$ is not empty.
\end{mylemma}

\begin{proof}
To state the conclusion of the lemma more concisely we define
flag $\Delta_C$ to be $*$ if $C$ is an empty set and $+$ otherwise.
The conclusion is ${\cal S} \models_{n}^{\Delta_{\cal C}} \varphi \ral \varphi'$.
The proof proceeds by induction on the proof tree showing
$\sequent[\cal C]{\cal A}{\varphi}{\varphi'}$
(keeping induction hypotheses universally quantified over $n$),
using an inner induction over $n$ in the \Circularity case.

\newcommand{\proofcase}[1]{\noindent\textbf{#1:}}

\proofcase{\Circularity}
The induction hypothesis states that for any $m \in \mathbb{N}$.
\begin{equation}\label{eq:circularityIH1}
\mbox{if }{\cal S} \models_{m}^{+} {\cal A}\mbox{ and }
{\cal S} \models_{m-1}^{+} {\cal C} \cup \{\varphi \ral \varphi'\}
\mbox{ then }{\cal S} \models_{m}^{+} \varphi \ral \varphi'.
\end{equation}
We will strengthen $\Delta_{\cal C}$ to ${+}$
and prove that for any $n \in \mathbb{N}$,
if ${\cal S} \models_{n}^{+} {\cal A}$ and
${\cal S} \models_{n - 1}^{+} {\cal C}$ then 
${\cal S} \models_{n}^{+} \varphi \ral \varphi'$.
Proceed by induction on $n$.
\begin{enumerate}
\item if $n = 0$, the conclusion
${\cal S} \models_0^{+} \varphi \ral \varphi'$ of the implication
is vacuously true.
\item for $n > 0$, we have an inner induction hypothesis that
${\cal S} \models_{n-1}^{+} {\cal A}$ and ${\cal S} \models_{n - 2}^{+} {\cal C}$
together imply ${\cal S} \models_{n-1}^{+} \varphi \ral \varphi'$,
and must prove that 
${\cal S} \models_{n}^{+} {\cal A}$ and ${\cal S} \models_{n - 1}^{+} {\cal C}$
together imply ${\cal S} \models_{n}^{+} \varphi \ral \varphi'$,
Instantiating the inner hypothesis with the stronger assumptions about
$\cal A$ and $\cal C$ gives ${\cal S} \models_{n-1}^{+} \varphi \ral \varphi'$.
With other assumptions this gives
${\cal S} \models_{n-1}^{+} {\cal C} \cup \{\varphi \ral \varphi'\}$.
Now we apply the outer induction hypothesis, Equation~\eqref{eq:circularityIH1},
with $m = n$ to conclude ${\cal S} \models_{n}^{+} \varphi \ral \varphi'$.
\end{enumerate}

This was the only case that made essential use of
the quantification over $n$ in the conclusion.
For the remaining cases, we assume that $n \in \mathbb{N}$ is already fixed,
and assumptions ${\cal S} \models_n^+ {\cal A}$ and $\models^+_{n - 1} {\cal C}$
are given in addition to the assumptions and inductive hypotheses coming from
the specific proof rule.
Inductive hypotheses with the same $\cal A$ and $\cal C$ as the conclusion
will be stated in the form
${\cal S} \models_n^{\Delta_{\cal C}} \varphi_a \ral \varphi_b$,
already specialized to $n$ and applied to the semantic assumptions on
$\cal A$ and $\cal C$.
Furthermore, the remaining cases all prove the conclusion
${\cal S} \models_n^+ \varphi \ral \varphi'$ directly from the definition,
so we also assume that a complete path $\tau = \gamma_1,\ldots,\gamma_k$
with $k \le n$ and a $\rho$ with $\mlmodels{\gamma_1}{\rho}{\varphi}$ are fixed.
This leaves the obligation to find an $i \in \{1,\ldots,k\}$
such that $\mlmodels{\gamma_i}{\rho}{\varphi'}$
and that $i > 1$ if $\cal C$ is nonempty.

\proofcase{\Transitivity}
We have inductive hypotheses that
${\cal S} \models_n^{\Delta_{\cal C}} \varphi \rightarrow \varphi_2$,
and for any $m \in \mathbb{N}$,
${\cal S} \models_m^+ {\cal A \cup C}$ implies
${\cal S} \models_m^* \varphi_2 \ra \varphi'$.
Applying the first inductive hypothesis to $\tau$ and $\rho$, we
receive an $i$ such that $\mlmodels{\gamma_i}{\rho}{\varphi_2}$.
Now we make separate cases on whether $i=1$.
\begin{itemize}
\item When $i = 1$, $\cal C$ must be empty, so our assumption
${\cal S} \models_n^+ {\cal A}$ suffices to instantiate the second
inductive hypothesis at $m=n$ to obtain
${\cal S} \models_n^* \varphi_2 \ra \varphi'$.
As $\mlmodels{\gamma_1}{\rho}{\varphi_2}$ we can apply this to
$\tau$ and $\rho$ to obtain a $j$ such that $\mlmodels{\gamma_j}{\rho}{\varphi'}$,
concluding this case.
\item Otherwise, the suffix of $\tau$ beginning with $\gamma_i$ has length
strictly less than $n$, so it suffices to instantiate the second
inductive hypothesis at $m=n-1$.
Our assumptions on $\cal C$ and $\cal A$ suffice to conclude
${\cal S} \models_{n-1}^+ {\cal A \cup C}$.
We obtain ${\cal S} \models_{n-1}^* \varphi_2 \ral \varphi'$.
Applying this to $\rho$ and the complete path $\gamma_i,\ldots,\gamma_k$
yields $j$ such that $\mlmodels{\gamma_{i+j-1}}{\rho}{\varphi'}$.
As $i+j-1 > 1$, we can conclude this case with $i+j-1$ whether or not $\cal C$ is empty.
\end{itemize}

\proofcase{\Step}
This rule has assumptions
$\models \varphi \ra \bigvee_{\lhs \Ra \rhs \ \in \ \Sem} \exists \FV{\lhs}\lhs$,
and for any $\lhs \Ra \rhs \in \Sem$
$\models \exists c \ (\varphi[c/\square] \andx \lhs[c/\square]) \andx \rhs \ra \varphi'$
.
By the first assumption $\gamma_1$ is not stuck, so complete path
$\tau$ has a second entry $\gamma_2$.
By the definition of a path, $\gamma_1 \RRs \gamma_2$.
By the definition of $\RRs$, there exists
$\varphi_l \Ra \varphi_r \in S$ and valuation $\rho'$ such that
$\mlmodels{\gamma_1}{\rho'}{\varphi_l}$ and
$\mlmodels{\gamma_2}{\rho'}{\varphi_r}$.
Let $X = \FV{\varphi, \varphi'}$ and $Y = \FV{\varphi_l, \varphi_r}$.
We assume without loss of generality that $X \cap Y = \emptyset$.
Fix any $\rho''$ which agrees with $\rho$ on $X$ and with $\rho'$ on $Y$.
With this choice $\mlmodels{\gamma_1}{\rho''}{\varphi}$, 
  $\mlmodels{\gamma_1}{\rho''}{\varphi_l}$, and
  $\mlmodels{\gamma_2}{\rho''}{\varphi_r}$.
  By the definition of satisfaction and the FOL translation, then also
  $\mlmodels{\gamma_2}{\rho''}
   {\exists c. (\varphi[c/\square] \andx \varphi_l[c/\square])}$
  and finally $\mlmodels{\gamma_2}{\rho''}
   {\exists c. (\varphi[c/\square] \andx \varphi_l[c/\square]) \andx \varphi_r}$.
  By the second assumption this implies $\mlmodels{\gamma_2}{\rho''}{\varphi'}$.
  As $\rho''$ agrees with $\rho$ on $X$, $\mlmodels{\gamma_2}{\rho}{\varphi'}$
  (and $2 > 1$), so taking $i=2$ concludes this case.

\proofcase{\Reflexivity}
This rule requires $\cal C$ be empty, and $\varphi = \varphi'$,
so $i=1$ suffices by $\mlmodels{\gamma_1}{\rho}{\varphi}$.

\proofcase{\Axiom}
This rule has assumption $\varphi \ral \varphi' \in \cal{A}$.
By the assumption on $\cal {A}$, ${\cal S} \models_{n}^{+} \varphi \ral \varphi'$.
Applying this to $\tau$ and $\rho$ gives an $i > 1$ 
with $\mlmodels{\gamma_i}{\rho}{\varphi'}$, concluding this case.

\proofcase{\Consequence}
We have inductive hypothesis that
${\cal S} \models_{n}^{\Delta_{\cal C}} \varphi_1 \ral \varphi_1'$,
and assumptions $\models \varphi \ra \varphi_1$
and $\models \varphi_1' \ra \varphi'$.
By the first implication $\mlmodels{\gamma_1}{\rho}{\varphi_1}$ as well,
so we can apply the inductive hypothesis with $\tau$ and $\rho$ to obtain
some $i$ such that $\mlmodels{\gamma_i}{\rho}{\varphi_1'}$
and $i > 1$ if $\cal C$ is nonempty.
By the second implication, $\mlmodels{\gamma_i}{\rho}{\varphi'}$ as well,
so this $i$ concludes this case.

\proofcase{\Case}
This rule requires $\varphi$ have the form $ \varphi_1 \orx \varphi_2$,
and we have inductive hypotheses
${\cal S} \models_n^{\Delta_{\cal C}} \varphi_1 \ral \varphi'$
and
${\cal S} \models_n^{\Delta_{\cal C}} \varphi_2 \ral \varphi'$.
Because $\mlmodels{\gamma_1}{\rho}{\varphi}$, either
$\mlmodels{\gamma_1}{\rho}{\varphi_1}$ or $\mlmodels{\gamma_1}{\rho}{\varphi_2}$.
In either case we can use the respective inductive hypotheses with
$\tau$ and $\rho$ to obtain an $i$ such that
$\mlmodels{\gamma_i}{\rho}{\varphi'}$ and $i > 1$ if $\cal C$ is nonempty,
concluding this case.

\proofcase{\Abstraction}
This rule requires $\varphi$ have the form $\exists X \varphi_1$
with $X \not\in \FV{\varphi'}$, and we
have the inductive hypothesis
${\cal S} \models_n^{\Delta_{\cal C}} \varphi_1 \ral \varphi'$.
Because $\mlmodels{\gamma_1}{\rho}{\exists X \varphi_1}$ there
exists a $\rho'$ which agrees with $\rho$ except on $X$ such that
$\mlmodels{\gamma_1}{\rho'}{\varphi_1}$.
Using the inductive hypothesis with $\tau$ and this $\rho'$
gives an $i$ such that $\mlmodels{\gamma_i}{\rho'}{\varphi'}$,
and $i > 1$ if $\cal C$ is nonempty.
As $X \not\in \FV{\varphi'}$, then also $\mlmodels{\gamma_i}{\rho}{\varphi'}$,
so this $i$ concludes this case.
\end{proof}

\section{Relative Completeness}\label{sec:completeness}

Here we show relative completeness: any valid all-path reachability property of
any program in any language with an operational semantics given as a
reachability system $\Sem$ is derivable with the proof system in
Fig.~\ref{fig:proof-system} from $\Sem$.
As with Hoare and dynamic logics, ``relative'' means we
assume an oracle capable of establishing validity in the first-order
theory of the state, which here is the configuration model ${\cal T}$.
Unlike the similar results for Hoare logics, which are separately proved for
each language taking into account the particularities of that language, we prove
relative completeness once and for all languages.
An immediate consequence of relative completeness
is that \Circularity is sufficient to derive any
repetitive behavior occurring in any program written in any language, and that
\Step is also sufficient to derive any non-deterministic behavior!

We establish the relative completeness under the
following assumptions:
\begin{center}
\framebox{
\begin{tabular}{@{ \ }l@{}}
\underline{\textsf{Framework:}} \\
The semantics reachability system $\Sem$ is \\
--- finite; \\
The configuration signature $\Sigma$ has \\
--- a sort $\mathbb{N}$; \\
--- constant symbols $0$ and $1$ of $\mathbb{N}$; \\
--- binary operation symbols $+$ and $\times$ on $\mathbb{N}$; \\
--- an operation symbol $\alpha : \Cfg \ra \mathbb{N}$. \\
The configuration model $\cal T$ interprets \\
--- $\mathbb{N}$ as the natural numbers; \\
--- constant and operation symbols on $\mathbb{N}$ as corresponding operations; \\
--- $\alpha : \Cfg \ra \mathbb{N}$ as an injective function.
\end{tabular}
}
\end{center}
The assumption that $\Sem$ is finite ensures \Step has a finite number of
prerequisites.
The assumption that the model ${\cal T}$ includes natural numbers with addition
and multiplication is a standard assumption (also made by Hoare and dynamic
logic completeness results) which allows the definition of G\"odel's $\beta$
predicate.
The assumption that the model ${\cal T}$ includes some injective function
$\alpha : \TCfg \ra \mathbb{N}$ (that is, the set of configurations
$\TCfg$ is countable) allows the encoding of a sequence of configurations into a
sequence of natural numbers.
We expect the operational semantics of any reasonable language to satisfy these
conditions.
Formally, we have the following
\begin{mytheorem}{Relative Completeness}
For any semantics $\Sem$ satisfying the assumptions above, if
$\Sem \models \varphi \Rra \varphi'$ then $\Sem \vdash \varphi \Rra
\varphi'$.
\end{mytheorem}

We present an informal proof sketch before going into the formal details,
Our proof relies on the fact that pattern reasoning in first-order matching
logic reduces to FOL reasoning in the model ${\cal T}$.
A key component of the proof is defining the $\coreach{\cvar}$ predicate in
plain FOL.
This predicate states that every complete $\RRs$-path $\tau$ starting at $\cvar$
includes some configuration satisfying $\varphi$.  
We express $\coreach{\cvar}$ using the auxiliary predicate
$\step{\cvar}{\cvar'}$
that encodes the one step transition relation ($\RRs$).
Fig.~\ref{fig:fol-encodings} shows both definitions.
As it is, $\coreach{\cvar}$ is not a proper FOL formula, as it
quantifies over a sequence of configurations.
This is addressed using the injective function $\alpha$ to encode universal
quantification over a sequence of configurations into universal quantification
over a sequence of integers, which is in turn encoded into quantification over
two integer variables using G\"odel's $\beta$ predicate (encoding shown
in Fig.~\ref{fig:fol-coreach}).

Next, using the definition above we encode the semantic validity of an all-path
reachability rule as FOL validity: $\Sem \models \varphi \Rra \varphi'$ iff
$\models \varphi \ra \coreach[\varphi']{\cvar}$.
Therefore, the theorem follows by \Consequence from the sequent
$\ssequent{\coreach[\varphi']{\cvar}}{\varphi'}$.
We derive this sequent by using \Circularity to add the rule to the
set of circularities, then by using \Step to derive one $\RRs$-step, and then by
using \Transitivity and \Axiom with the rule itself to derive the remaining
$\RRs$-steps (circularities can be used after \Transitivity).
The formal derivation uses all eight proof rules. 

Also recall that, as discussed in Section~\ref{sec:matching-logic}, matching
logic is a methodological fragment of the FOL theory of the model ${\cal T}$.
For technical convenience, in this section we work with the FOL translations
$\varphi^\square$ instead of the matching logic formulae $\varphi$.
We mention that in all the formulae used in this section, $\square$ only occurs
in the context $\square = t$, thus we stay inside the methodological fragment.
For the duration of the proof, we let $\cvar, \cvar', \cvar[0], \ldots,
\cvar[n]$ be distinct variables of sort $\Cfg$ which do not appear free in the
rules in $\Sem$).
We also let $\gamma, \gamma', \gamma_0, \ldots, \gamma_n$ range over (not
necessarily distinct) configurations in the model ${\cal T}$, that is, over
elements in $\TCfg$, and let $\rho, \rho'$ range over valuations $\Var \ra {\cal
T}$.

\subsection{Encoding Transition System Operations in FOL}\label{sec:rc-fol}

\begin{figure}[t]
\[\begin{array}{@{}r@{ \ \ }c@{ \ \ }l@{}}
\step{\cvar}{\cvar'} & \equiv & \displaystyle \bigvee_{\mu \equiv \lhs \Ra \rhs
\in \Sem} \exists \FV{\mu} \ (\squaresubst{\lhs}{\cvar} \andx
\squaresubst{\rhs}{\cvar'}) \\
\coreach{\cvar} & \equiv & \displaystyle \forall n \forall \cvar[0] \ldots
\cvar[n] \Bigg(\square = \cvar[0] \ra \!\!\bigwedge\limits_{0 \leq i < n}\!\!
\step{\cvar[i]}{\cvar[i + 1]} \ra \neg\exists \cvar[n + 1] \
\step{\cvar[n]}{\cvar[n + 1]} \ra \!\!\bigvee\limits_{0 \leq i \leq n}\!\!
\squaresubst{\varphi}{\cvar[i]}\Bigg)
\end{array}\]
\caption{\label{fig:fol-encodings}FOL encoding of one step transition relation
and all-path reachability.}
\end{figure}

Fig.~\ref{fig:fol-encodings} shows the definition of the one step transition
relation ($\RRs$) and of the configurations that reach $\varphi$ on all and
complete paths.
The former is a (proper) FOL formula, while the later is not, as it quantifies
over a sequence of configuration.
In Section~\ref{sec:rc-godel} we use G\"odel's $\beta$ predicate to define
$\folcoreach$, a FOL formula equivalent to $\coreach{\cvar}$.

First, we establish the following general purpose lemma
\begin{mylemma}\label{lem:square-subst}
$\mlmodels{\rho(\cvar)}{\rho}{\varphi^\square}$ iff $\rho \models
\squaresubst{\varphi^\square}{\cvar}$.
\end{mylemma}
\begin{proof}
With the notation in Definition~\ref{dfn:ml2fol},
$\mlmodels{\rho(c)}{\rho}{\varphi^\square}$ iff $\rho^{\rho(c)} \models
\varphi^\square$.
Notice that if a valuation agrees on two variables, then it satisfies a formula
iff it satisfies the formula obtained by substituting one of the two variables
for the other.
In particular, since $\rho^{\rho(c)}(\square) = \rho^{\rho(c)}(c)$, it follows
that $\rho^{\rho(c)} \models \varphi^\square$ iff $\rho^{\rho(c)} \models
\varphi^\square[c/\square]$.
We notice that $\square$ does not occur in $\varphi^\square[c/\square]$, thus
$\rho^{\rho(c)} \models \varphi^\square[c/\square]$ iff $\rho \models
\varphi^\square[c/\square]$, and we are done.
\end{proof}

The following lemma states that $\step{\cvar}{\cvar'}$ actually has the semantic
properties its name suggests.
\begin{mylemma}\label{lem:step-model}
$\rho \models \step{\cvar}{\cvar'}$ iff $\rho(\cvar) \RRs \rho(\cvar')$.
\end{mylemma}
\begin{proof}
Assume $\rho \models \step{\cvar}{\cvar'}$.
Then, by the definition of $\step{\cvar}{\cvar'}$, there exists some rule $\mu
\equiv \lhs \Ra \rhs \in \Sem$ such that $\rho \models \exists \FV{\mu} \
(\lhs[\cvar/\square] \andx \rhs[\cvar'/\square])$.
Further, since $\cvar$ and $\cvar'$ do not occur in $\mu$, there exists some
$\rho'$ which agrees with $\rho$ on $\cvar$ and $\cvar'$ such that $\rho'
\models \lhs[\cvar/\square]$ and $\rho' \models \rhs[\cvar'/\square]$.
By Lemma~\ref{lem:square-subst}, $\rho' \models \lhs[c/\square]$ iff
$\mlmodels{\rho'(\cvar)}{\rho'}{\lhs}$ and $\rho' \models \rhs[\cvar'/\square]$
iff $\mlmodels{\rho'(\cvar')}{\rho'}{\rhs}$, so
$\mlmodels{\rho'(\cvar)}{\rho'}{\lhs}$ and
$\mlmodels{\rho'(\cvar')}{\rho'}{\rhs}$.
Since $\rho$ and $\rho'$ agree on $\cvar$ and $\cvar'$, it follows that
$\mlmodels{\rho(c)}{\rhoS}{\lhs}$ and $\mlmodels{\rho(c')}{\rhoS}{\rhs}$.
By Definition~\ref{dfn:mlr}, we conclude $\rho(\cvar) \RRs \rho(\cvar')$.

Conversely, assume $\rho(\cvar) \RRs \rho(\cvar')$.
Then, by Definition~\ref{dfn:mlr}, there exist some rule $\mu \equiv \lhs \Ra
\rhs \in \Sem$ and some $\rho'$ for which $\mlmodels{\rho(\cvar)}{\rho'}{\lhs}$
and $\mlmodels{\rho(\cvar')}{\rho'}{\rhs}$.
Further, since $\cvar$ and $\cvar'$ do not occur in $\mu$, we can choose $\rho'$
to agree with $\rho$ on $\cvar$ and $\cvar'$.
Hence, $\mlmodels{\rho'(\cvar)}{\rho'}{\lhs}$ and
$\mlmodels{\rho'(\cvar')}{\rho'}{\rhs}$.
By Lemma~\ref{lem:square-subst}, $\mlmodels{\rho'(\cvar)}{\rho'}{\lhs}$ iff
$\rho' \models \lhs[\cvar/\square]$ and $\mlmodels{\rho'(\cvar')}{\rho'}{\rhs}$
iff $\rho' \models \rhs[\cvar'/\square]$, so $\rho' \models \lhs[\cvar/\square]$
and $\rho' \models \rhs[\cvar'/\square]$.
Since the free variables occurring in $\lhs[\cvar/\square] \andx
\rhs[\cvar'/\square]$ are $\FV{\mu} \cup \{ \cvar, \cvar' \}$ and $\rho$ and
$\rho'$ agree on $\cvar$ and $\cvar'$, it follows that $\rho \models \exists
\FV{\mu} \ (\lhs[\cvar/\square] \andx \rhs[\cvar'/\square])$.
By the definition of $\step{\cvar}{\cvar'}$, we conclude $\rho \models
\step{\cvar}{\cvar'}$.
\end{proof}

The following lemma introduces a formula encoding a complete path of fixed
length.
\begin{mylemma}\label{lem:coreach-model-partial}
$\rho \models \bigwedge\limits_{0 \leq i < n} \step{\cvar[i]}{\cvar[i + 1]}
\andx \not\exists \cvar[n + 1] \ \step{\cvar[n]}{\cvar[n + 1]}$ iff
$\rho(\cvar[0]), \ldots, \rho(\cvar[n + 1])$ is a complete $\RRs$-path.
\end{mylemma}
\begin{proof}
By Lemma~\ref{lem:step-model}, we have that $\rho(\cvar[i]) \RRs \rho(\cvar[i +
1])$ iff $\rho' \models \step{\cvar[i]}{\cvar[i + 1]}$, for each $0 \leq i < n$.
Further, $\rho(\cvar[0]), \ldots, \rho(\cvar[n + 1])$ is complete, iff there
does not exist $\gamma$ such that $\rho(\cvar[n]) \RRs \gamma$.
Again, by Lemma~\ref{lem:step-model}, that is iff $\rho \models \not\exists
\cvar[n + 1] \ \step{\cvar[n]}{\cvar[n + 1]}$.
We conclude that $\rho \models \bigwedge\limits_{0 \leq i < n}
\step{\cvar[i]}{\cvar[i + 1]} \andx \not\exists \cvar[n + 1] \
\step{\cvar[n]}{\cvar[n + 1]}$ iff $\rho(\cvar[0]), \ldots, \rho(\cvar[n + 1])$
is a complete $\RRs$-path, and we are done.
\end{proof}

The following lemma states that $\coreach{\cvar}$ actually has the semantic
properties its name suggests.
\begin{mylemma}\label{lem:coreach-model}
$\mlmodels{\gamma}{\rho}{\coreach{\cvar}}$ iff for all complete $\RRs$-paths
$\tau$ starting with $\gamma$ it is the case that
$\mlmodels{\gamma'}{\rho}{\varphi}$ for some $\gamma' \in \tau$.
\end{mylemma}
\begin{proof}
First we prove the direct implication.
Assume $\mlmodels{\gamma}{\rho}{\coreach{\cvar}}$, and let $\tau \equiv
\gamma_0, \ldots, \gamma_n$ be a complete $\RRs$-path starting with $\gamma$.
Then let $\rho'$ agree with $\rho$ on $\FV{\varphi}$ such that $\rho'(n) = n$
and $\rho'(\cvar[i]) = \gamma_i$ for each $0 \leq i \leq n$.
According to the definition of $\coreach{\cvar}$, we have that
\[\mlmodels{\gamma}{\rho'}{\square = \cvar[0] \andx \bigwedge\limits_{0 \leq i <
n} \step{\cvar[i]}{\cvar[i + 1]} \andx \not\exists \cvar[n + 1] \
\step{\cvar[n]}{\cvar[n + 1]} \ra \bigvee\limits_{0 \leq i \leq n}
\squaresubst{\varphi}{\cvar[i]}}\]
Since, $\gamma = \gamma_0$ and $\rho'(\cvar[0]) = \gamma_0$, it follows that
$\rho' \models \square = \cvar[0]$.
Further, by Lemma~\ref{lem:coreach-model-partial}, since $\rho'(\cvar[0]),
\ldots, \rho'(\cvar[n])$ is a complete $\RRs$-path, it must be the case that
\[\rho' \models \bigwedge\limits_{0 \leq i < n} \step{\cvar[i]}{\cvar[i + 1]}
\andx \not\exists \cvar[n + 1] \ \step{\cvar[n]}{\cvar[n + 1]}\]
Thus, as $\square$ does not occur in any $\squaresubst{\varphi}{\cvar[i]}$, we
conclude that $\rho' \models \bigvee\limits_{0 \leq i \leq n}
\squaresubst{\varphi}{\cvar[i]}$, that is, $\rho' \models
\squaresubst{\varphi}{\cvar[i]}$ for some $0 \leq i \leq n$.
By Lemma~\ref{lem:square-subst}, $\rho' \models \squaresubst{\varphi}{\cvar[i]}$
iff $\mlmodels{\gamma_i}{\rho'}{\varphi}$.
Since $\rho$ agrees with $\rho'$ on $\FV{\varphi}$, we conclude that
$\mlmodels{\gamma_i}{\rho}{\varphi}$.

Conversely, assume that if $\tau$ is a finite and complete $\RRs$-path starting
with $\gamma$.
Then $\mlmodels{\gamma'}{\rho}{\varphi}$ for some $\gamma' \in \tau$.
Let $\rho'$ agree with $\rho$ on $\FV{\varphi}$.
Then we prove that
\[\mlmodels{\gamma}{\rho'}{\square = \cvar[0] \andx \bigwedge\limits_{0
\leq i < n} \step{\cvar[i]}{\cvar[i + 1]} \andx \not\exists \cvar[n + 1] \
\step{\cvar[n]}{\cvar[n + 1]} \ra \bigvee\limits_{0 \leq i \leq n}
\squaresubst{\varphi}{\cvar[i]}}\]
Specifically, assume $\mlmodels{\gamma}{\rho'}{\square = \cvar[0] \andx
\bigwedge\limits_{0 \leq i < n} \step{\cvar[i]}{\cvar[i + 1]} \andx \not\exists
\cvar[n + 1] \ \step{\cvar[n]}{\cvar[n + 1]}}$.
As $\square$ does not occur in any $\cvar{\cvar[i]}{\cvar[i + 1]}$, by
Lemma~\ref{lem:coreach-model-partial}, it follows that $\rho'(\cvar[0]), \ldots,
\rho'(\cvar[n])$ is a complete $\RRs$-path.
Further, $\mlmodels{\gamma}{\rho'}{\square = \cvar}$, implies that
$\rho'(\cvar[0]), \ldots, \rho'(\cvar[n])$ starts with $\gamma$.
Thus, there exists some $0 \leq i \leq n$ such that
$\mlmodels{\rho'(\cvar[i])}{\rho}{\varphi}$, or equivalently, since $\rho$ and
$\rho'$ agree on $\FV{\varphi}$, such that
$\mlmodels{\rho'(\cvar[i])}{\rho'}{\varphi}$.
By Lemma~\ref{lem:square-subst}, $\mlmodels{\rho'(\cvar[i])}{\rho'}{\varphi}$
iff $\rho' \models \squaresubst{\varphi}{\cvar[i]}$.
Therefore, we have that $\mlmodels{\gamma}{\rho'}{\bigvee\limits_{0 \leq i \leq
n} \squaresubst{\varphi}{\cvar[i]}}$.
Finally, since $\rho'$ is an arbitrary valuation which agrees with $\rho$ on
$\FV{\varphi}$, by the definition of $\coreach{\cvar}$ we can conclude that
$\mlmodels{\gamma}{\rho}{\coreach{\cvar}}$, and we are done.
\end{proof}

The following lemma establishes a useful property of $\coreach{\cvar}$.
\begin{mylemma}\label{lem:coreach-csq}
\[\models \coreach{\cvar} \ra \varphi \orx (\exists \cvar' \
\step{\cvar}{\cvar'} \andx \forall \cvar' \ (\step{\cvar}{\cvar'} \ra
\squaresubst{\coreach{\cvar'}}{\cvar'}))\]
\end{mylemma}
\begin{proof}
We prove that if $\mlmodels{\gamma}{\rho}{\coreach{\cvar}}$ then
\[\mlmodels{\gamma}{\rho}{\varphi \orx (\exists \cvar' \ \step{\cvar}{\cvar'}
\andx \forall \cvar' \ (\step{\cvar}{\cvar'} \ra
\squaresubst{\coreach{\cvar'}}{\cvar'}))}\]
By Lemma~\ref{lem:coreach-model}, we have that for all complete $\RRs$-paths
$\tau$ starting with $\gamma$ it is the case that
$\mlmodels{\gamma''}{\rho}{\varphi}$ for some $\gamma'' \in \tau$.
We distinguish two cases
\begin{itemize}

\item $\mlmodels{\gamma}{\rho}{\varphi}$. We are trivially done.

\item $\notmlmodels{\gamma}{\rho}{\varphi}$.
Then $\gamma$ must have $\RRs$-successors.
Indeed, assume the contrary.
Then $\tau \equiv \gamma$ is a complete $\RRs$-path.
It follows that $\mlmodels{\gamma}{\rho}{\varphi}$, which is a contradiction.
Thus, there exists some $\gamma'$ such that $\gamma \RRs \gamma'$.
By Lemma~\ref{lem:step-model}, that is iff $\rho \models \exists \cvar' \
\step{\cvar}{\cvar'}$.
Further, let $\gamma'$ be a $\RRs$-successor of $\gamma$ and $\tau'$ a complete
$\RRs$-path starting with $\gamma'$.
Then, $\gamma\tau$ is a complete $\RRs$-path starting with $\gamma$.
Thus, there exists some $\gamma'' \in \gamma\tau'$ such that
$\mlmodels{\gamma''}{\rho}{\varphi}$.
Since $\notmlmodels{\gamma}{\rho}{\varphi}$, it follows that $\gamma'' \in
\tau'$.
Notice that $\gamma'$ is an arbitrary configuration and $\tau'$ an arbitrary
$\RRs$-path, therefore by Lemma~\ref{lem:coreach-model} and
Lemma~\ref{lem:square-subst}, we can conclude that $\rho \models \forall \cvar'
\ (\step{\cvar}{\cvar'} \ra \squaresubst{\coreach{\cvar'}}{\cvar'}))$.
\qedhere
\end{itemize}
\end{proof}

\subsection{Formula G\"odelization}\label{sec:rc-godel}

\begin{figure}
\begin{small}
\[\begin{array}{rcl}
\folcoreach & \equiv & \forall n \forall a \forall b \ (\exists \cvar \
(\gbeta{a}{b}{0}{\alpha(\cvar)} \andx \square = \cvar) \\
& & \hspace*{2.35ex} \andx \forall i \ (0 \leq i \andx i < n \ra \exists \cvar
\exists \cvar' \ (\gbeta{a}{b}{i}{\alpha(\cvar)} \andx \gbeta{a}{b}{i +
1}{\alpha(\cvar')} \andx \step{\cvar}{\cvar'})) \\
& & \hspace*{2.35ex} \andx \exists \cvar \ (\gbeta{a}{b}{n}{\alpha(\cvar)} \andx
\not\exists \cvar' \ \step{\cvar}{\cvar'}) \\
& & \hspace*{2.35ex} \ra \exists i \ (0 \leq i \andx i \leq n \andx \exists
\cvar \ (\gbeta{a}{b}{i}{\alpha(\cvar)} \andx \squaresubst{\varphi}{\cvar}))) 
\end{array}\]
\caption{\label{fig:fol-coreach}FOL definition of $\coreach{\cvar}$}
\end{small}
\end{figure}

Fig.~\ref{fig:fol-coreach} defines $\folcoreach$, the FOL equivalent of
$\coreach{\cvar}$ using G\"odel's $\beta$ predicate. Formally, we have the
following
\begin{mylemma}
$\models \coreach{\cvar} \leftrightarrow \folcoreach$.
\end{mylemma}
\begin{proof}
Let us choose some arbitrary but fixed values for $n$, $a$ and $b$, and let
$\rho'$ such that $\rho$ and $\rho'$ agree on $\Var \setminus \{ n, a, b \}$ and
$\rho'(n) = n$ and $\rho'(a) = a$ and $\rho'(b) = b$.
According to the definition of the $\beta$ predicate, there exists a unique
integer sequence $j_0, \ldots, j_n$ such that $\gbeta{a}{b}{i}{j_i}$ holds for
each $0 \leq i \leq n$.
Since $\alpha$ is injective, we distinguish two cases
\begin{itemize}

\item there exists some $0 \leq i \leq n$ such that there is not any $\gamma_i$
with $\alpha(\gamma_i) = j_i$

\item there exists a unique sequence $\gamma_0, \ldots, \gamma_n$ such that
$\alpha(\gamma_i) = j_i$ for each $0 \leq i \leq n$.

\end{itemize}
In the former case, if $i = n$ we get that $\rho' \not\models \exists \cvar \
(\gbeta{a}{b}{n}{\alpha(\cvar)} \andx \not\exists \cvar' \ \step{\cvar}{\cvar'}$
while if $0 \leq i < n$ we get that $\rho' \not\models \exists \cvar \exists
\cvar' \ (\gbeta{a}{b}{i}{\alpha(\cvar)} \andx \gbeta{a}{b}{i +
1}{\alpha(\cvar')} \andx \step{\cvar}{\cvar'})$ as in both cases we can not pick
a value for $\cvar$.
Thus, $(\gamma, \rho')$ does not satisfy left-hand-side of the implication in
$\folcoreach$, and we conclude that $(\gamma, \rho')$ satisfies the implication.

In the later case, we have that there is a unique way of instantiating the
existentially quantified variables $\cvar$ and $\cvar'$ in each sub-formula in
which they appear, as they are always arguments of the $\beta$ predicate.
Thus, $\mlmodels{\gamma}{\rho'}{\exists \cvar \ (\gbeta{a}{b}{0}{\alpha(\cvar)}
\andx \square = \cvar)}$ iff $\gamma = \gamma_0$.
By Lemma~\ref{lem:coreach-model-partial}, we have that
\[\begin{array}{rcl}
\rho' & \models & \forall i \ (0 \leq i \andx i < n \ra \exists \cvar \exists
\cvar' \ (\gbeta{a}{b}{i}{\alpha(\cvar)} \andx \gbeta{a}{b}{i +
1}{\alpha(\cvar')} \andx \step{\cvar}{\cvar'})) \\
& & \andx \exists \cvar \ (\gbeta{a}{b}{n}{\alpha(\cvar)} \andx \not\exists
\cvar' \ \step{\cvar}{\cvar'})
\end{array}\]
iff $\gamma_0\ldots\gamma_n$ is a complete $\RRs$-path.
Finally, by Lemma~\ref{lem:square-subst}
\[\rho' \models \exists i \ (0 \leq i \andx i \leq n \andx \exists \cvar
\ (\gbeta{a}{b}{i}{\alpha(\cvar)} \andx \squaresubst{\varphi}{\cvar}))\]
iff $\mlmodels{\gamma_i}{\rho'}{\varphi}$ for some $0 \leq i \leq n$.

We conclude that $(\gamma, \rho')$ satisfies the implication in $\folcoreach$
iff
\begin{itemize}

\item there is no sequence $\gamma_0, \ldots, \gamma_n$ such that
$\alpha(\gamma_i) = j_i$ for each $0 \leq i \leq n$ 

\item the unique sequence $\gamma_0, \ldots, \gamma_n$ such that
$\alpha(\gamma_i) = j_i$ for each $0 \leq i \leq n$ is either not starting at
$\gamma$, not a complete $\RRs$-path or contains some $\gamma'$ such that
$\mlmodels{\gamma'}{\rho}{\varphi}$, as $\rho$ and $\rho'$ agree on $\Var
\setminus \{ n, a, b \}$.

\end{itemize}

According to the property of $\beta$, for each sequence $j_0, \ldots, j_n$ there
exist some values for $a$ and $b$.
Since $n$, $a$ and $b$ are chosen arbitrary, we conclude that
$\mlmodels{\gamma}{\rho}{\folcoreach}$ iff for all complete $\RRs$-paths $\tau$
starting at $\gamma$, there exists some $\gamma' \in \tau$ such that
$\mlmodels{\gamma'}{\rho}{\varphi}$.
By Lemma~\ref{lem:coreach-model}, we have that the above iff
$\mlmodels{\gamma}{\rho}{\coreach{\cvar}}$, and we are done.
\end{proof}

\subsection{Encoding Semantic Validity in FOL}

Now we can use the \textit{coreach} predicate to encode the semantic validity of
a rule $\varphi \Rra \varphi'$ in FOL.
Formally
\begin{mylemma}\label{lem:semantic-validity}
If $\Sem \models \varphi \Rra \varphi'$ then $\models \varphi \ra
\coreach[\varphi']{}$.
\end{mylemma}
\begin{proof}
Follows from the definition of semantic validity of $\varphi \Rra \varphi'$ and
Lemma~\ref{lem:coreach-model}.
\end{proof}

\subsection{Relative Completeness}\label{sec:rc-rc}

A matching logic formula $\psi$ is {\em patternless} iff $\square$ does not
occur in $\psi$.
Then we have the following lemma stating that we can derive one step on all paths
\begin{mylemma}\label{lem:step}
$\sequent[\cal C]{\cal A}{\square = \cvar \andx \exists \cvar' \
\step{\cvar}{\cvar'} \andx \psi}{\exists \cvar' \ (\square = \cvar' \andx
\step{\cvar}{\cvar'}) \andx \psi}$ where $\psi$ is a patternless formula.
\end{mylemma}
\begin{proof}
We derive the rule by applying the \Step proof rule with the following
prerequisites
\[\models \square = \cvar \andx \exists \cvar' \ \step{\cvar}{\cvar'} \andx \psi
\ra \bigvee_{\lhs \Ra \rhs \in \Sem} \exists \FV{\lhs} \ \lhs\]
and for each $\lhs \Ra \rhs \ \in \ \Sem$ (since $\square$ does not occur in
$\psi$)
\[\models \exists \cvar'' \ (\cvar'' = \cvar \andx \exists \cvar' \
\step{\cvar}{\cvar'} \andx \squaresubst{\lhs}{\cvar''}) \andx \rhs \andx \psi
\ra \exists \cvar' \ (\square = \cvar' \andx \step{\cvar}{\cvar'}) \andx \psi\]

\noindent
For the first prerequisite, we have the following (using the definition of
$\step{\cvar}{\cvar'}$)
\[\begin{array}{cl}
& \square = \cvar \andx \exists \cvar' \ \step{\cvar}{\cvar'} \andx \psi \\ 
\ra & \square = \cvar \andx \exists \cvar' \ \step{\cvar}{\cvar'} \\
\leftrightarrow & \displaystyle \square = \cvar \andx \exists \cvar' \
\bigvee_{\mu \equiv \lhs \Ra \rhs \in \Sem} \exists \FV{\mu} \
(\squaresubst{\lhs}{\cvar} \andx \squaresubst{\rhs}{\cvar'}) \\
\ra & \displaystyle \square = \cvar \andx \exists \cvar' \ \bigvee_{\mu \equiv
\lhs \Ra \rhs \in \Sem} \exists \FV{\mu} \ \squaresubst{\lhs}{\cvar} \\
\ra & \displaystyle \square = \cvar \andx \bigvee_{\mu \equiv \lhs \Ra \rhs \in
\Sem} \exists \FV{\lhs} \ \squaresubst{\lhs}{\cvar} \\
\ra & \displaystyle \bigvee_{\mu \equiv \lhs \Ra \rhs \in \Sem} \exists
\FV{\lhs} \ \lhs \\
\end{array}\]

\noindent
For the second prerequisite, let $\lhs \Ra \rhs \in \Sem$. Then we have that
\[\begin{array}{cl}
& \exists \cvar'' \ (\cvar'' = \cvar \andx \exists \cvar' \ \step{\cvar}{\cvar'}
\andx \squaresubst{\lhs}{\cvar''}) \andx \rhs \andx \psi \\
\ra & \squaresubst{\lhs}{\cvar} \andx \rhs \andx \psi \\
\ra & \exists \cvar' \ (\square = \cvar' \andx \squaresubst{\lhs}{\cvar} \andx
\squaresubst{\rhs}{\cvar'}) \andx \psi \\
\ra & \displaystyle \exists \cvar' \ (\square = \cvar' \andx \bigvee_{\mu \equiv
\lhs \Ra \rhs \in \Sem} (\squaresubst{\lhs}{\cvar} \andx
\squaresubst{\rhs}{\cvar'})) \andx \psi \\
\ra & \displaystyle \exists \cvar' \ (\square = \cvar' \andx \bigvee_{\mu \equiv
\lhs \Ra \rhs \in \Sem} \exists \FV{\mu} \ (\squaresubst{\lhs}{\cvar} \andx
\squaresubst{\rhs}{\cvar'}) \andx \psi \\
\ra & \exists \cvar' \ (\square = \cvar' \andx \step{\cvar}{\cvar'}) \andx \psi
\\ 
\end{array}\]
and we are done.
\end{proof}

The following three lemmas show that we can derive a rule stating that all the
configurations reaching $\varphi$ in the transition system actually reach
$\varphi$.  
\begin{mylemma}\label{lem:coreach-partial}
If
\[\sequent{\cal A}{\square = \cvar \andx \exists \cvar' \ \step{\cvar}{\cvar'}
\andx \forall \cvar' \ (\step{\cvar}{\cvar'} \ra
\squaresubst{\coreach{\cvar'}}{\cvar'})}{\varphi}\]
then $\sequent{\cal A}{\coreach{\cvar}}{\varphi}$.
\end{mylemma}
\begin{proof}
By Lemma~\ref{lem:coreach-csq}
\[\models \coreach{\cvar} \leftrightarrow \varphi \orx (\exists \cvar' \
\step{\cvar}{\cvar'} \andx \forall \cvar' \ (\step{\cvar}{\cvar'} \ra
\squaresubst{\coreach{\cvar'}}{\cvar'}))\]
Thus, by \Consequence and \Case, it suffices to derive
\[\begin{array}{l}
\sequent{\cal A}{\varphi}{\varphi} \\
\sequent{\cal A}{\square = \cvar \andx \exists \cvar' \ \step{\cvar}{\cvar'}
\andx \forall \cvar' \ (\step{\cvar}{\cvar'} \ra
\squaresubst{\coreach{\cvar'}}{\cvar'})}{\varphi} \\
\end{array}\]
The first sequent follows by \Reflexivity.
The second sequent is part of the hypothesis, and we are done.
\end{proof}

\begin{mylemma}\label{lem:coreach-step}
\[\sequent{\cal A}{\square = \cvar \andx \exists \cvar' \ \step{\cvar}{\cvar'}
\andx \forall \cvar' \ (\step{\cvar}{\cvar'} \ra
\squaresubst{\coreach{\cvar'}}{\cvar'})}{\varphi}\]
\end{mylemma}
\begin{proof}
Let $\mu$ be the rule we want to derive, namely
\[\square = \cvar \andx \exists \cvar' \ \step{\cvar}{\cvar'} \andx \forall
\cvar' \ (\step{\cvar}{\cvar'} \ra \squaresubst{\coreach{\cvar'}}{\cvar'}) \Rra
\varphi\]
Then $\rlsequent{\cal A}{\mu}$ follows by \Circularity from $\rlsequent[\{ \mu
\}]{\cal A}{\mu}$.
Hence, by \Transitivity, it suffices to derive the two sequents below
\[\begin{array}{l}
\sequent[\{ \mu\}]{\cal A}{\square = \cvar \andx \exists \cvar' \
\step{\cvar}{\cvar'} \andx \forall \cvar' \ (\step{\cvar}{\cvar'} \ra
\squaresubst{\coreach{\cvar'}}{\cvar'})}{\varphi'} \\
\sequent{{\cal A} \cup \{ \mu \}}{\varphi'}{\varphi} \\
\end{array}\]
where $\varphi' \equiv \exists \cvar' \ (\square = \cvar' \andx
\step{\cvar}{\cvar'}) \andx \forall \cvar' \ (\step{\cvar}{\cvar'} \ra
\squaresubst{\coreach{\cvar'}}{\cvar'})$.
The first sequent follows by Lemma~\ref{lem:step} with $\psi \equiv \forall
\cvar' \ (\step{\cvar}{\cvar'} \ra \squaresubst{\coreach{\cvar'}}{\cvar'})$.
For the second sequent, by \Abstraction with $\{ \cvar' \}$ and \Consequence
with
\[\models \square = \cvar' \andx \step{\cvar}{\cvar'} \andx \forall \cvar' \
(\step{\cvar}{\cvar'} \ra \squaresubst{\coreach{\cvar'}}{\cvar'}) \ra
\coreach{\cvar'}\]
it suffices to derive $\sequent{{\cal A} \cup \{ \mu
\}}{\coreach{\cvar'}}{\varphi}$.
Then, by Lemma~\ref{lem:coreach-partial}, we are left to derive
\[\sequent{{\cal A} \cup \{ \mu \}}{\square = \cvar \andx \exists \cvar' \
\step{\cvar}{\cvar'} \andx \forall \cvar' \ (\step{\cvar}{\cvar'} \ra
\squaresubst{\coreach{\cvar'}}{\cvar'})}{\varphi}\]
that is, $\rlsequent{{\cal A} \cup \{ \mu \}}{\mu}$, which trivially follows by
\Axiom and we are done.
\end{proof}

\begin{mylemma}\label{lem:coreach}
$\sequent{\cal A}{\coreach{\cvar}}{\varphi}$.
\end{mylemma}
\begin{proof}
By Lemma~\ref{lem:coreach-partial}, it suffices to derive
\[\sequent{\cal A}{\square = \cvar \andx \exists \cvar' \ \step{\cvar}{\cvar'}
\andx \forall \cvar' \ (\step{\cvar}{\cvar'} \ra \coreach{\cvar'})}{\varphi}\]
which follows by Lemma~\ref{lem:coreach-step}.
\end{proof}

Finally, we can establish the main result.
Assume $\Sem \models \varphi \Rra \varphi'$.
By Lemma~\ref{lem:semantic-validity}, we have that $\models \varphi \ra
\coreach[\varphi']{}$.
Further, by Lemma~\ref{lem:coreach}, we have that $\Sem \vdash
\coreach[\varphi']{\cvar} \Rra \varphi'$.
Then, by \Consequence, it follows that $\Sem \vdash \varphi \Rra \varphi'$.


\section{Related Work}

Using Hoare logic~\cite{hoare-69} to prove concurrent programs correct dates
back to Owicki and Gries~\cite{owicki-gries-1}.
In the rely-guarantee method proposed by Jones~\cite{rely-guarantee}
each thread relies on some properties being satisfied by the other threads, and
in its turn, offers some guarantees on which the other threads can rely.
O'Hearn~\cite{1236121} advances a Separation Hypothesis in the context of
separation logic~\cite{reynolds-02} to achieve compositionality: the state can
be partitioned into separate portions for each process and relevant resources,
respectively, satisfying certain invariants.
More recent research focuses on improvements over both of the above methods and
even combinations of them (e.g.,~\cite{local-rg,rg-sl,csl,csl-1}).  


The satisfaction of all-path-reachability rules can also be understood
intuitively in the context of temporal logics.
Matching logic formulae can be thought of as state formulae, and
reachability rules as temporal formulae.
Assuming $\text{CTL}^\ast$ on finite traces, the semantics rule
$\varphi \Ra \varphi'$ can be expressed as
$\varphi \implies E \bigcirc \varphi'$, while an all-path reachability
rule $\varphi \Rra \varphi'$ can be expressed as $\varphi \implies A
\Diamond \varphi'$.
However, unlike in $\text{CTL}^\ast$, the $\varphi$ and $\varphi'$ formulae
of reachability rules $\varphi \Ra \varphi'$ or $\varphi \Rra \varphi'$
share their free variables.
Thus, existing proof systems for temporal logics (e.g., the $\text{CTL}^\ast$
one by Pnueli and Kesten~\cite{pnueli-kesten}) are not directly comparable with
our approach.
In general, $\text{CTL}^\ast$ formulae only have atomic predicates, while in
our approach we consider only a specific temporal structure ($\varphi \implies
A \Diamond \varphi'$), but the matching logic formulae $\varphi$ and $\varphi'$
can express more complex state properties.

This approach has been extended to handle safety properties of distributed
system, specified as rewrite theories~\cite{DBLP:conf/lopstr/SkeirikSM17}.
Specifically, it can verify that $\varphi$ is an invariant, or a co-invariant,
of a non-terminating distributed system by checking that $\varphi$ holds after
each application of the \Step proof rule. The soundness of the approach is
proved by transforming the underlying transition system from non-terminating to
terminating by adding a transition from each state to a newly-introduced end
state (where $\varphi$ does not hold).



Bae et al.~\cite{DBLP:conf/rta/BaeEM13}, Rocha and
Meseguer~\cite{DBLP:conf/calco/RochaM11}, and Rocha et al.~\cite{rocha-2014-wrla}
use narrowing to perform symbolic reachability analysis in a transition system
associated to a unconditional rewrite theory for the purposes of verification.
There are two main differences between their work and ours.
First, they express state predicates in equational theories. Matching logic is
more general, being first-order logic over a model of configurations ${\cal T}$.
Consequently, the \Step proof rule takes these issues into account when
considering the successors of a state.
Second, they use rewrite systems for symbolic model checking.
Our work is complementary, in the sense that we use the operational semantics
for program verification, and check properties more similar to those in Hoare
logic.

Dynamic logic~\cite{Harel84dynamiclogic} adds modal operators to FOL to embed
program fragments within specifications.
For example, $\psi\rightarrow[{\tt s}]\psi'$ means ``after executing {\tt s} in
a state satisfying $\psi$, a state may be reached which satisfies $\psi'$''.
KeY~\cite{KeYBook2007} offers automatic verification for Java based on dynamic
logic.
Dynamic logic has been extended to support reasoning about cyber-physical
systems, where the transitions may have a continuous or probabilistic
component~\cite{DBLP:conf/sbmf/MadeiraNMB14}.
Matching logic also combines programs and specifications for static properties,
but dynamic properties are expressed in reachability logic which has a
language-independent proof system that works with any operational semantics,
while dynamic logic still requires language-specific proof rules (e.g.,
invariant rules for loops).

\paragraph{Language-independent proof systems}
A first proof system is introduced in~\cite{rosu-stefanescu-2012-icalp},
while~\cite{rosu-stefanescu-2012-fm} presents a mechanical translation from Hoare
logic proof derivations for \IMP into derivations in the proof system.
The \Circularity proof rule is introduced in~\cite{rosu-stefanescu-2012-oopsla}.
Finally, \cite{rosu-stefanescu-ciobaca-moore-2013-lics} supports operational
semantics given with conditional rules, like small-step and big-step. 
All these previous results can only be applied to deterministic programs.
The coinductive nature of this proof system was clarified in%
~\cite{moore-pena-rosu-2018-esop}, which presents a coinduction principle
that captures the essence of the \Circularity rule, and which is sufficient
for program verification when used within a system such as ZFC or Coq that
already supports basic logical reasoning and working with collections
of configurations.

\section{Conclusion and Future Work}

This paper introduces a sound and (relatively) complete language-independent
proof system which derives program properties holding along all execution paths
(capturing partial correctness for non-deterministic programs), directly from an
operational semantics.
The proof system separates reasoning about deterministic language features
(via the operational semantics) from reasoning about
non-determinism (via the proof system).
Thus, all we need in order to verify programs in a language is an operational
semantics for the respective language.

We believe that existing techniques such as rely-guarantee and concurrent
separation logic could be used in conjunction with our proof system to achieve
semantically grounded and compositional verification.

Our approach handles operational semantics given with unconditional rules, like
in the \K framework, PLT-Redex, and CHAM, but it cannot handle operational semantics
given with conditional rules, like big-step and small-step (rules with
premises).
Extending the presented results to work with conditional rules would boil down
to extending the \Step proof rule, which derives the fact that $\varphi$ reaches
$\varphi'$ in one step along all execution paths. 
Such an extended \Step would have as prerequisites whether the left-hand side of
a semantics rule matches (like the existing \Step) and additionally whether its
premises hold.
The second part would require an encoding of reachability in first-order logic,
which is non-trivial and most likely would result in a first-order logic over
a richer model than $\cal T$.
The difficulty arises from the fact that \Step must ensure all successors of
$\varphi$ are in $\varphi'$.
Thus, this extension is left as future work.

\paragraph{Acknowledgements}
We would like to thank the anonymous reviewers and the FSL members for their
valuable feedback on an early version of this paper.
The work presented in this paper was supported in part by the Boeing grant on
"Formal Analysis Tools for Cyber Security" 2014-2015, the NSF grant CCF-1218605,
the NSA grant H98230-10-C-0294, the DARPA HACMS program as SRI subcontract
19-000222, the Rockwell Collins contract 4504813093, and the (Romanian)
SMIS-CSNR 602-12516 contract no. 161/15.06.2010.

\bibliographystyle{splncs}
\bibliography{refs}

\end{document}